\documentclass[journal]{IEEEtran}
\usepackage{amsmath}
\usepackage{graphicx, graphics, theorem, times, amsfonts, amsmath, amssymb, cite, caption}
\usepackage{tikz}
\usetikzlibrary{shapes,arrows}
\tikzstyle{phantom vertex} = [ ellipse, 
                               anchor = center, 
                               minimum height = 1*\unit, 
                               minimum width  = 1*\unit,
                               inner sep=0pt,
                               anchor=center]
\tikzstyle{red vertex}   = [black, fill = red!20,   phantom vertex, draw]
\tikzstyle{black vertex} = [black, fill = black!20, phantom vertex, draw]
\tikzstyle{blue vertex}  = [black, fill = blue!20,  phantom vertex, draw]
\tikzstyle{green vertex} = [black, fill = green!20,  phantom vertex, draw]
\tikzstyle{vertex}       = [draw, phantom vertex]

\tikzstyle{point} = [ellipse, inner sep=0pt, draw, fill=white, anchor = center,
                     minimum height = 0.05*\unit, minimum width  = 0.05*\unit]
\usepackage{pgfplots}
\usepackage{color}
\usepackage{needspace}

\usepackage{hyperref}
\usepackage{multirow}
\usepackage{rotating}
\usepackage{flushend, cleveref, subcaption}
\usepackage{bm}
\input{mysymbol.sty}
\newtheorem{lemma}{\hspace{0pt}\bf Lemma}
\newtheorem{claim}{\hspace{0pt}\bf Claim}
\newtheorem{proposition}{\hspace{0pt}\bf Proposition}

\newtheorem{remark}{\hspace{0pt}\bf Remark}

\newtheorem{definition}{\hspace{0pt}\bf Definition}

\newcommand{\QED}{\hfill\ensuremath{\blacksquare}}
\begin{document}
\title{Stability and Continuity of Centrality Measures in Weighted Graphs}
\author{Santiago Segarra and Alejandro Ribeiro 

\thanks{Work supported by NSF CCF-1217963. The authors are with the Department of Electrical and Systems Engineering, University of Pennsylvania, 200 South 33rd Street, Philadelphia, PA 19104. Email: \{ssegarra, aribeiro\}@seas.upenn.edu. Part of the results in this paper appeared in \cite{cSegarraRibeiro14}.}}

\maketitle

% A B S T R A C T %%%%%%%%%%%%%%%%%%%%%%%%%%%%%%%%%%%%%%%%%%%%%%%%%%%%%%%%%%%%
\begin{abstract} This paper presents a formal definition of stability for node centrality measures in weighted graphs. It is shown that the commonly used measures of degree, closeness and eigenvector centrality are stable whereas betweenness centrality is not. An alternative definition of the latter that preserves the same centrality notion while satisfying the stability criteria is introduced. Continuity is presented as a less stringent alternative to stability. Betweenness centrality is shown to be not only unstable but discontinuous. Numerical experiments in synthetic random networks and real-world data show that, in practice, stability and continuity imply different levels of robustness in the presence of noisy data. In particular, the stable betweenness centrality is shown to exhibit resilience against noise that is absent in the discontinuous and unstable standard betweenness centrality, while preserving a similar notion of centrality.
\end{abstract}
% E N D : A B S T R A C T %%%%%%%%%%%%%%%%%%%%%%%%%%%%%%%%%%%%%%%%%%%%%%%%%%%%%%%%

%%%%%%%%%%%%%%%%%%%%%%%%%%%%%%%%%%%%%%%%%%%%%%%%%%%%%%%%%%%%%%%%%%%
%%%%%%%%%%%%%%%%%%%%%%   I   N   T   R   O   D   U   C   T   I   O   N    %%%%%%%%%%%%%%%%%%%%%%%%%%
%%%%%%%%%%%%%%%%%%%%%%%%%%%%%%%%%%%%%%%%%%%%%%%%%%%%%%%%%%%%%%%%%%%
\section{Introduction}\label{sec:introduction}

In any graph or network, the topology determines an influence structure among the nodes or agents. Peripheral nodes have limited impact on the dynamics of the network whereas central nodes have a major effect on the behavior of the whole graph. Identifying the most important nodes in a network helps in explaining the network's dynamics, e.g.  the distribution of power in exchange networks \cite{Cooketal83} or migration in biological networks \cite{Garrowayetal08}, as well as in designing optimal ways to externally influence the network, e.g. attack vulnerability of networks \cite{Holmeetal02}. Node centrality measures are tools designed to identify such important agents. However, node importance is a rather vague concept and can be interpreted in various ways, giving rise to multiple coexisting centrality measures, the most common being degree \cite{Shaw54, Nieminen73}, closeness \cite{Sabidussi66, Beauchamp65}, eigenvector \cite{Bonacich72}, and betweenness \cite{Freeman77} centrality. In degree centrality, the importance or centrality of a node is measured by the number of nodes it can immediately influence, i.e., its neighborhood. In closeness centrality, importance is measured in terms of how fast information can travel from a given node to every other node in the network. In eigenvector centrality, a refinement of degree centrality, the importance of a node is computed as a function of the importance of its neighbors. Finally, in betweenness centrality, the centrality of a node is given by the frequency of this node belonging to the shortest path between other two nodes in the network. 

The ability of a centrality measure to be robust to noise in the network data is of practical importance. In the past decade, stability has been used as a parameter to compare the performance of different centrality measures \cite{CostenbaderValente03, Borgattietal06, ZemljiHlebec05}. In these papers, an empirical approach was followed by comparing stability indicators measured in both random and real-world networks for different centrality measures. However, no formal theory was developed explaining the different behaviors among measures. Our first contribution is a formal definition of stability and continuity of centrality measures. We also show that all frequently used measures are stable and continuous with the exception of betweenness centrality. In addition, we propose an alternative definition of betweenness centrality which is stable. Finally, through numerical experiments in synthetic and real-word networks, we demonstrate that stability and continuity are different and important properties, and show that the alternative definition of betweenness centrality behaves better than the standard betweenness centrality while preserving a similar notion of centrality.

Stability is formally defined in Section \ref{sec_node_centrality_and_stability}. In order to build such definition, we need to rely on a metric on the space of weighted graphs with a common node and edge set. In Sections \ref{sec_degree_centrality} to \ref{sec_eigenvector_centrality}, we analyze the stability of the most frequently used centrality measures and in Section \ref{sec_stable_centrality_measure} we propose an alternative definition of betweenness centrality that guarantees stability while maintaining the same concept of node centrality. The concept of continuity as a milder requirement for robustness is introduced in Section \ref{sec_stability_and_continuity}. In Section \ref{sec_numerical_experiments}, we illustrate how our formal definitions of stability and continuity are correlated with practical robustness indicators by analyzing the behavior of all the common centrality measures as well as the stable betweenness centrality proposed in random networks and two real-world networks: the network of air traffic between airports in United States and the network of interactions between sectors of the United States economy.

%%%%%%%%%%%%%%%%%%%%%%%%%%%%%%%%%%%%%%%%%%%%%%%%%%%%%%%%%%%%%%%%%%%

%%%%%%%%%%%%%%%%%%%%%%%%%%%%%%%%%%%%%%%%%%%%%%%%%%%%%%%%%%%%%%%%%%%
%%%%%%%%%%%%%%%%%%%%%%   P   R   E   L   I   M   I   N   A   R   I   E   S  %%%%%%%%%%%%%%%%%%%%%%%%%
%%%%%%%%%%%%%%%%%%%%%%%%%%%%%%%%%%%%%%%%%%%%%%%%%%%%%%%%%%%%%%%%%%%
\section{Preliminaries}
\label{sec_preliminaries}

In the present paper we consider weighted and directed graphs or networks. Formally, we define a graph $G=(V, E, W)$ as a triplet formed by a finite set of $n$ nodes or vertices $V$, a set of directed edges $E \subset V \times V$ where $(x, y) \in E$ represents an edge from $x \in V$ to $y \in V$, and a set of positive weights $W: E \to \reals_{++}$ defined on each edge. The weights can be associated to similarities between nodes, i.e. the higher the weight the more similar the nodes are, or dissimilarities, depending on the application. The graphs considered here do not contain self-loops, i.e., $(x, x) \not\in E$ for all $x \in V$. For any given sets $V$ and $E$, denote by $\ccalG_{(V,E)}$ the space of all graphs with $V$ as node set and $E$ as edge set. This implies that two graphs $G, H \in \ccalG_{(V,E)}$ can only differ in their weights. 

An alternative representation of a graph is through its adjacency matrix $A \in \reals^{n \times n}$. If there exists an edge from node $i$ to node $j$, then $A_{ij}$ takes the value of the corresponding weight. Otherwise, $A_{ij}$ is null. Requiring graphs not to contain self-loops is equivalent to requiring the diagonal of $A$ to consist of all zeros. Observe that if two graphs $G, H \in \ccalG_{(V,E)}$, then the null entries of the corresponding adjacency matrices must coincide.

In the definition of centrality measures, the concepts of path and path length are important. Given a graph $(V, E, W)$ and $x, x' \in V$, a path $P(x, x')$ is an \emph{ordered} sequence of nodes in $V$, 
\begin{equation}\label{eqn_definition_path}
   P(x, x')=[x=x_0, x_1, \ldots , x_{l-1}, x_l=x'],
\end{equation}
which starts at $x$ and finishes at $x'$ and $e_i=(x_i, x_{i+1}) \in E$ for $i=0, \ldots, l-1$. We say that $P(x, x')$ links or connects $x$ to $x'$. 
The links $e_i$ of a path are the edges connecting consecutive nodes of the path in the direction given by the path. Specifically when $W$ is associated to dissimilarities, we define the length of a given path $P(x, x')=[x=x_0,\ldots, x_l=x']$ as the sum of the weights $\sum_{i=0}^{l-1} W(e_i)$ encountered when traversing its links in order. Given the graph $G=(V,E,W)$, we define the shortest path function $s_G: V \times V \to \overline{\reals}_+$ where the shortest path length $s_G(x, x')$ between nodes $x, x' \in V$ is defined as
\begin{equation}\label{eqn_shortest_path}
s_G(x,x') := \min_{P(x, x')} \,\, \sum_{i=0}^{l-1} W(x_i, x_{i+1}).
\end{equation}
Whenever there is no possible path linking $x$ to $x'$ in a graph $G$, we say that $s_G(x,x') = \infty$.
%%%%%%%%%%%%%%%%%%%%%%%%%%%%%%%%%%%%%%%%%%%%%%%%%%%%%%%%%%%%%%%%%%%%%

%%%%%%%%%%%%%%%%%%%%%%%%%%%%%%%%%%%%%%%%%%%%%%%%%%%%%%%%%%%%%%%%%%%
%%%%%%%%%%   N   O    D    E      C   E   N   T   R   A   L   I   T   Y      A   N   D      S   T   A   B   I   L   I   T   Y   %%%%%%%%%%%%%% 
%%%%%%%%%%%%%%%%%%%%%%%%%%%%%%%%%%%%%%%%%%%%%%%%%%%%%%%%%%%%%%%%%%%
\section{Node Centrality and Stability}
\label{sec_node_centrality_and_stability}

Node centrality is a measure of the importance of a node within a graph. This importance is based on the location of the node within the graph and not on the intrinsic nature of this node. More precisely, given a graph $(V, E, W)$, a centrality measure $C: V \to \reals_+$ assigns a nonnegative centrality value to every node such that the higher the value the more central the node is. The centrality ranking imposed by $C$ on the node set $V$ is in general more relevant than the absolute centrality values. Very often, this centrality ranking relies on an underlying characteristic of the nodes. E.g., airports which are hubs for some airline have high centrality in an air transportation network. In this way, centrality detects fundamental roles played by nodes within the graph. Ideally, this detection should be invariant to small perturbations in the edge weights.

To formalize this notion of robustness against perturbations, we define the metric $d_{(V,E)}: \ccalG_{(V,E)} \times \ccalG_{(V,E)} \to \reals_+$ on the space of graphs $\ccalG_{(V,E)}$ containing $V$ as node set and $E$ as edge set, as follows
\begin{equation}\label{eqn_metric_definition}
d_{(V,E)}(G, H) :=\sum_{e \in E} |W(e)-W'(e)| = \sum_{i,j} |A_{ij}-A'_{ij}|,
\end{equation}
where $G=(V, E, W)$ and $H=(V, E, W')$, and have adjacency matrices $A$ and $A'$, respectively. To see that $d_{(V,E)}$ is a well-defined metric, notice that it computes the $\ell_1$ distance between two vectors obtained by stacking the values in $W$ and $W'$. The metric $d_{(V,E)}$ enables the formal definition of stability presented next.
%

%%%%%%%%%%%%%%%%%%%%%%%%  D E F I N I T I O N :      S T A B I L I T Y  %%%%%%%%%%%%%%%%%%%%%%%%%%%%%
\begin{definition}\label{def_stability}
A centrality measure $C$ is \emph{stable} if, for every vertex set $V$, edge set $E$ and any two graphs $G, H \in \ccalG_{(V,E)}$, 
\begin{equation}\label{eqn_stability_definition}
\left| C^G(x) - C^H(x) \right| \leq K_G \, d_{(V,E)}(G,H), 
\end{equation}
for every node $x \in V$, where $K_G$ is a constant for every graph $G$, $C^G(x)$ is the centrality value of node $x$ in graph $G$ and similarly for $H$.
\end{definition}
%%%%%%%%%%%%%%%%%%%%%%%%%%%%%%%%%%%%%%%%%%%%%%%%%%%%%%%%%%%%%%%%%%%%%%%%

The above definition states that a centrality measure is stable if the difference in centrality values for a given node in two different graphs is bounded by a constant $K_G$ times the distance between these graphs. The constant $K_G$ only depends on graph $G$ and must be valid for every graph $H$ to which $G$ is being compared. Moreover, the inclusion of $K_G$ in \eqref{eqn_stability_definition} ensures that the stability of a centrality measure does not depend on the appearance of a normalization term in the definition of the measure. In particular, if graph $H$ is a perturbed version of $G$, any stable centrality measure ensures that the change in centrality due to this perturbation is bounded for every node. This generates a robust measure in the presence of noise as we illustrate through examples in Section \ref{sec_numerical_experiments}. In the following sections we analyze the stability of the most frequently used centrality measures.

%%%%%%%%%%%%%%%%%%%%%%%%%%%%%%%%%%%%%%%%%%%%%%%%%%%%%%%%%%%%%%%%%%%%%

\subsection{Degree centrality}\label{sec_degree_centrality}

Degree centrality is a local measure of the importance of a node within a graph. The degree centrality measure $C_D$ of a node $x$ in an undirected weighted graph $(V, E, W)$ is given by the sum of the weights of the edges incident to node $x$, that is,
\begin{equation}\label{def_degree_centrality}
C_D(x) := \sum_{x' | (x, x') \in E} W(x, x').
\end{equation}
For directed graphs, degree centrality is usually unfolded into two different measures: in-degree and out-degree centrality. The out-degree centrality $C_{OD}$ measure is computed as in \eqref{def_degree_centrality}, whereas the in-degree centrality $C_{ID}$ is computed as follows
\begin{equation}\label{def_in_degree_centrality}
C_{ID}(x) := \sum_{x' | (x', x) \in E} W(x', x).
\end{equation}
The degree centrality measure is applied to graphs where the weights in $W$ represent similarities between the nodes. In this way, a high degree centrality value of a given node means that this node has a large number of neighbors and is closely connected to them. Although the degree centrality measure has a number of limitations related to its locality \cite{Borgatti05}, it is stable as we state next.

%%%%%%%%%%%%%%%%%%%%%%%%%%%%%%% P R O P O S I T I O N %%%%%%%%%%%%%%%%%%%%%%%%%%%%%%
\begin{proposition}\label{prop_degree_is_stable}
The degree $C_D$, out-degree $C_{OD}$ and in-degree $C_{ID}$ centrality measures in \eqref{def_degree_centrality} and \eqref{def_in_degree_centrality} are stable as defined in Definition \ref{def_stability} with $K_G = 1$.
\end{proposition}
%%%%%%%%%%%%%%%%%%%%%%%%%%%%%%%%% P R O O F %%%%%%%%%%%%%%%%%%%%%%%%%%%%%%%%%
\begin{myproofnoname}
Consider two arbitrary graphs in $\ccalG_{(V,E)}$, $G = (V, E, W)$ and $H = (V, E, W')$. From the definition of degree centrality \eqref{def_degree_centrality}, we obtain
\begin{align}\label{eqn_proof_degree_is_stable_010}
|C_D^G(x)&-C_D^{H}(x)| \\
& = \left| \sum_{x' | (x, x') \in E} W(x, x') -  \sum_{x' | (x, x') \in E} W'(x, x') \right|. \nonumber
\end{align}
Consolidating the summations in \eqref{eqn_proof_degree_is_stable_010} and applying the triangular inequality we obtain that
\begin{align}\label{eqn_proof_degree_is_stable_020}
|C_D^G(x)&-C_D^{H}(x)|  \leq \sum_{x' | (x, x') \in E}  \left| W(x, x') - W'(x, x') \right|. 
\end{align}
By summing the right hand side of \eqref{eqn_proof_degree_is_stable_020} over all edges instead of just a subset of them we obtain
\begin{align}\label{eqn_proof_degree_is_stable_030}
|C_D^G(x)-C_D^{H}(x)| \leq \sum_{e \in E}  \left| W(e) - W'(e) \right|. 
\end{align}
The right hand side of \eqref{eqn_proof_degree_is_stable_030} is exactly $d_{(V,E)}(G, H)$ [cf. \eqref{eqn_metric_definition}], showing inequality \eqref{eqn_stability_definition} for $K_G=1$. When considering directed graphs, the proof can be replicated to show that the in-degree $C_{ID}$  and out-degree $C_{OD}$ centrality measures are stable.
\end{myproofnoname}
%%%%%%%%%%%%%%%%%%%%%%%%%%%%%%%%%%%%%%%%%%%%%%%%%%%%%%%%%%%%%%%%%%%%%%%

A consequence of the stability property of degree centrality shown in Proposition \ref{prop_degree_is_stable} is the limited effect that a perturbation in the weights of a graph has on the centrality values. In Section \ref{sec_numerical_experiments}, we illustrate this in both synthetic and real-world networks.

\subsection{Closeness centrality}\label{sec_closeness_centrality}

Closeness is a relevant centrality measure when we are interested in how fast information can spread from one node to every other node in a network. The most commonly used definition of closeness centrality is the one in \cite{Sabidussi66} where the centrality $C_C(x)$ of a node $x$ in a graph $G=(V, E, W)$ is defined as the inverse of the sum of the shortest path lengths from this node to every other node in the graph, i.e.
\begin{equation}\label{def_closeness_centrality_2}
C_C(x) := \left( \sum_{x' \in V} s_G(x, x') \right)^{-1}.
\end{equation}
For \eqref{def_closeness_centrality_2} to make sense, the weights in $W$ must represent dissimilarities between the nodes. Moreover, in general, we consider strongly connected graphs so that every shortest path has finite length. This implies that \eqref{def_closeness_centrality_2} is well-defined. However, as done in \cite{Freeman78}, we will work with the \emph{decentrality} version $\bar{C}_C$, where the lower the value the more central the node, defined as
\begin{equation}\label{def_closeness_centrality}
\bar{C}_C(x) := \sum_{x' \in V} s_G(x, x').
\end{equation}
Since we are ultimately interested in the centrality ranking being impervious to perturbations, it is immediate that the ranking stability of $C_C$ and of $\bar{C}_C$ are equivalent since they are related by a strictly decreasing function. In the following proposition, we show stability of closeness decentrality.

%%%%%%%%%%%%%%%%%%%%%%%%%%%%%%% P R O P O S I T I O N %%%%%%%%%%%%%%%%%%%%%%%%%%%%%%
\begin{proposition}\label{prop_closeness_is_stable}
The closeness decentrality measure $\bar{C}_C$ in \eqref{def_closeness_centrality} is stable as defined in Definition \ref{def_stability} with $K_G = n$.
\end{proposition}

In proving Proposition \ref{prop_closeness_is_stable}, we use the following lemma which upper bounds the difference between a shortest path in two different graphs by the distance between these graphs.
%%%%%%%%%%%%%%%%%%%%%%%%%%%%%%%%% L E M M A %%%%%%%%%%%%%%%%%%%%%%%%%%%%%%%%%
\begin{lemma}\label{lem_shortest_path_less_distance}
Given two graphs $G=(V, E, W)$ and $H=(V, E, W')$ then, for every pair of nodes $x, x' \in V$ such that $x$ is connected to $x'$,  
\begin{equation}\label{eqn_shortest_path_less_distance}
\left| s_G(x, x') - s_{H}(x, x') \right| \leq d_{(V,E)}(G,H),
\end{equation}
where $s_G$ and $s_{H}$ are the shortest path lengths defined in \eqref{eqn_shortest_path}. 
\end{lemma}
%%%%%%%%%%%%%%%%%%%%%%%%%%%%%%%%% P R O O F %%%%%%%%%%%%%%%%%%%%%%%%%%%%%%%%%
\begin{myproofnoname}
Consider two arbitrary graphs $G = (V, E, W)$ and $H=(V, E, W')$ and two connected nodes $x, x' \in V$. Suppose that one shortest path from $x$ to $x'$ in $G$ is given by the path $P(x, x')=[x=x_0, x_1, \ldots, x_l=x']$ and one shortest path from $x$ to $x'$ in $H$ is given by $P'(x, x')=[x=x'_0, x'_1, \ldots, x'_{l'}=x']$. Then, by the definition of shortest path in \eqref{eqn_shortest_path}, we have that
\begin{align}\label{eqn_proof_lem_shortest_path_less_distance_010}
|s_G(x, x')  -   s_{H}&(x, x')|  = \\
&  \left| \sum_{i=0}^{l-1} W(x_i, x_{i+1})  -  \sum_{i=0}^{l'-1} W'(x'_i, x'_{i+1})  \right|  . \nonumber
\end{align}
Without loss of generality, assume that the graph $G$ has a larger shortest path, i.e., $s_G(x, x') \geq s_{H}(x, x')$. By this assumption, the difference between the shortest path lengths in \eqref{eqn_proof_lem_shortest_path_less_distance_010} is nonnegative even without the absolute value. Hence, if instead of considering the shortest path $P(x, x')$ in $G$ we consider a (possibly) different path such as $P'(x, x')$, we can assure that the difference between these two lengths is not going to decrease. Then, it follows that
\begin{align}\label{eqn_proof_lem_shortest_path_less_distance_020}
|s_G(x, x')   \! - \! s_{H}(x, x')|  \! \leq \! \left| \sum_{i=0}^{l'-1} W(x'_i, x'_{i+1}) - W'(x'_i, x'_{i+1}) \right| \!.
\end{align}
A direct application of the triangular inequality yields
\begin{align}\label{eqn_proof_lem_shortest_path_less_distance_030}
|s_G(x, x') \! - \! s_{H}(x, x')| \! \leq \! \sum_{i=0}^{l'-1} \left|W(x'_i, x'_{i+1}) - W'(x'_i, x'_{i+1}) \right| \!.
\end{align}
Finally, if on the right hand side of \eqref{eqn_proof_lem_shortest_path_less_distance_030} instead of summing over the links in $P'(x, x')$ we sum over all edges in the set $E$, we obtain another upper bound given by
\begin{align}\label{eqn_proof_lem_shortest_path_less_distance_040}
|s_G(x, x') \! - \! s_{H}(x, x')| \leq \sum_{e \in E} \left|W(e) - W'(e) \right|.
\end{align}
The right hand side of \eqref{eqn_proof_lem_shortest_path_less_distance_040} is exactly $d_{(V,E)}(G, H)$ [cf. \eqref{eqn_metric_definition}], concluding the proof.
\end{myproofnoname}

We can now leverage Lemma \ref{lem_shortest_path_less_distance} to show the stability of closeness decentrality.

%%%%%%%%%%%%%%%%%%%%%%%%%% P R O O F   O F   P R O P O S I T I O N %%%%%%%%%%%%%%%%%%%%%%%%%%%
\begin{myproof}[of Proposition \ref{prop_closeness_is_stable}]
Given two strongly connected graphs $G=(V,E,W)$ and $H=(V,E,W')$, from the definition of $\bar{C}_C$ in \eqref{def_closeness_centrality} we have that
\begin{align}\label{eqn_proof_closeness_is_stable_010}
|\bar{C}_C^G(x) - \bar{C}_C^H(x)| = \left| \sum_{x' \in V} s_G(x, x') -  \sum_{x' \in V} s_H(x, x') \right|. 
\end{align}
Consolidating the summations in \eqref{eqn_proof_closeness_is_stable_010} and applying the triangular inequality we obtain that
\begin{align}\label{eqn_proof_closeness_is_stable_020}
|\bar{C}_C^G(x) - \bar{C}_C^H(x)| \leq \sum_{x' \in V}  \left| s_G(x, x') - s_H(x, x') \right|.
\end{align}
Using the result in Lemma \ref{lem_shortest_path_less_distance} we conclude that
\begin{align}\label{eqn_proof_closeness_is_stable_030}
|\bar{C}_C^G(x) - \bar{C}_C^H(x)| \leq \sum_{x' \in V}  d_{(V,E)}(G, H) = n \,\, d_{(V,E)}(G, H), 
\end{align}
showing inequality \eqref{eqn_stability_definition} for $K_G = n$.
\end{myproof}
%%%%%%%%%%%%%%%%%%%%%%%%%%%%%%%%%%%%%%%%%%%%%%%%%%%%%%%%%%%%%%%%%%%%%%%%

Some alternative definitions of closeness centrality exist \cite{MoxleyMoxley74, Bavelas50} including that in \cite{Beauchamp65} where the measure in \eqref{def_closeness_centrality} is normalized by $n-1$. However, since normalization constants can be absorbed into $K_G$, stability does not depend on the appearance of normalization terms. 

\begin{remark}\label{rem_closeness_centrality}\normalfont
If we adopt the convention that $\infty - \infty = 0$, then the result in Lemma \ref{lem_shortest_path_less_distance} is true even when nodes $x$ and $x'$ are not connected. This, in turn, implies that Proposition \ref{prop_closeness_is_stable} can be shown for general graphs and the requirement of strong connectivity can be dropped.
\end{remark}

\subsection{Betweenness centrality}\label{sec_betweenness_centrality}

Centrality can be interpreted as the possibility of a node to control the communication or the optimal flow within a graph. Betweenness centrality takes this position by giving higher centrality values to nodes that fall within the shortest path of many pairs of nodes. Formally, given a graph $G=(V, E, W)$ and three arbitrary nodes $x, x', x'' \in V$, denote by $\sigma_{x' x''}$ the number of shortest paths from $x'$ to $x''$, i.e. the number of paths $P(x', x'')$ of length $s_G(x', x'')$, and by $\sigma_{x' x''}(x)$ the number of these shortest paths that go through node $x$. For convenience, we define $\sigma_{x x}=1$ for all $x \in V$. Notice that since $G$ might be directed, we can have that $\sigma_{x'x''} \neq \sigma_{x''x'}$ for some $x', x'' \in V$. 
The betweenness centrality $C_B(x)$ for any given node $x \in V$ is defined as \cite{Freeman77}
\begin{equation}\label{eqn_betweenness_centrality}
C_B(x) :=\sum_{\substack{x', x'' \in V \\ x' \neq x \neq x'' }} \frac{\sigma_{x' x''}(x)}{\sigma_{x' x''}}.
\end{equation}
In \eqref{eqn_betweenness_centrality}, we compute the betweenness centrality value of a node $x \in V$ by sequentially looking at the shortest paths between any two nodes distinct from $x$ and summing the proportion of shortest paths that contain node $x$. As was the case for closeness centrality, the weights in $W$ should denote dissimilarities for $C_B$ to be a reasonable measure of centrality. Sometimes \eqref{eqn_betweenness_centrality} is normalized by the number of pairs in the network or the maximum centrality value achievable  \cite{Freeman77, WhiteBorgatti94, Brandes08} such that $C_B(x)$ takes values in the interval $[0, 1]$. However, we are interested in comparing centrality values between different nodes within a network and these comparisons are invariant to any normalization. Moreover, the stability property does not depend on normalizing constants since these can get absorbed by $K_G$ [cf. \eqref{eqn_stability_definition}]. Hence, we omit the normalizing constant in definition \eqref{eqn_betweenness_centrality}.

Despite its extensive use in the study of both technological \cite{Onnelaetal07} and social \cite{Newman01} networks, the betweenness centrality measure is not stable as we show next.

%%%%%%%%%%%%%%%%%%%%%%%%%%%   F   I   G   U   R   E   %%%%%%%%%%%%%%%%%%%%%%%%%%%%%%%%
\begin{figure*}
\centering
\centerline{\def \thisplotscale {0.65}
\def \unit {\thisplotscale cm}

{\small
\begin{tikzpicture}[-stealth, shorten >=2, scale = \thisplotscale]

%%% FIRST NETWORK %%%%%%%%%
% NAME
    \node at (0-3-3.3,0.8) {$G$};
    
% NODES

    \node [blue vertex, minimum height=0.6cm, minimum width=0.6cm] at (0-3,0) (1) {$x_1$};
    \node [blue vertex, minimum height=0.6cm, minimum width=0.6cm] at (0-3,-2) (2) {$x_2$};    
    \node [blue vertex, minimum height=0.6cm, minimum width=0.6cm] at (-3-3,-1) (3) {$x_3$};
    \node [blue vertex, minimum height=0.6cm, minimum width=0.6cm] at (3-3,-1) (4) {$x_4$};
    \node [blue vertex, minimum height=0.6cm, minimum width=0.6cm] at (-4-4,0.5) (5) {$x_5$};
    \node [blue vertex, minimum height=0.6cm, minimum width=0.6cm] at (-4-4,-2.5) (6) {$x_6$};
    \node [blue vertex, minimum height=0.6cm, minimum width=0.6cm] at (4-2,0.5) (7) {$x_7$};
    \node [blue vertex, minimum height=0.6cm, minimum width=0.6cm] at (4-2,-2.5) (8) {$x_8$};

%EDGES
    
    \draw[latex'-latex'] (1) -- node [pos=0.3, above left] {$1$} (3);
    \draw[latex'-latex'] (1) -- node [pos=0.3, above right] {$1$} (4);
    \draw[latex'-latex'] (3) -- node [below] {$1$} (2);
    \draw[latex'-latex'] (4) -- node [below] {$1$} (2);
    \draw[latex'-latex'] (5) -- node [below left, pos=0.35] {$1$} (3);
    \draw[latex'-latex'] (6) -- node [above left, pos=0.35] {$1$} (3);
    \draw[latex'-latex'] (7) -- node [below right, pos=0.35] {$1$} (4);
    \draw[latex'-latex'] (8) -- node [above right, pos=0.35] {$1$} (4);

    %%% SECOND NETWORK %%%%%%%%%
% NAME
    \node at (7.7,0.8) {$H$};
    
% NODES

    \node [blue vertex, minimum height=0.6cm, minimum width=0.6cm] at (11,0) (1p) {$x_1$};
    \node [blue vertex, minimum height=0.6cm, minimum width=0.6cm] at (11,-2) (2p) {$x_2$};    
    \node [blue vertex, minimum height=0.6cm, minimum width=0.6cm] at (8,-1) (3p) {$x_3$};
    \node [blue vertex, minimum height=0.6cm, minimum width=0.6cm] at (14,-1) (4p) {$x_4$};
    \node [blue vertex, minimum height=0.6cm, minimum width=0.6cm] at (7-1,0.5) (5p) {$x_5$};
    \node [blue vertex, minimum height=0.6cm, minimum width=0.6cm] at (7-1,-2.5) (6p) {$x_6$};
    \node [blue vertex, minimum height=0.6cm, minimum width=0.6cm] at (15+1,0.5) (7p) {$x_7$};
    \node [blue vertex, minimum height=0.6cm, minimum width=0.6cm] at (15+1,-2.5) (8p) {$x_8$};

%EDGES
    
    \draw[latex'-latex'] (1p) -- node [pos=0.3, above left] {$1+\epsilon$} (3p);
    \draw[latex'-latex'] (1p) -- node [pos=0.3, above right] {$1+\epsilon$} (4p);
    \draw[latex'-latex'] (3p) -- node [below] {$1$} (2p);
    \draw[latex'-latex'] (4p) -- node [below] {$1$} (2p);
    \draw[latex'-latex'] (5p) -- node [below left, pos=0.35] {$1$} (3p);
    \draw[latex'-latex'] (6p) -- node [above left, pos=0.35] {$1$} (3p);
    \draw[latex'-latex'] (7p) -- node [below right, pos=0.35] {$1$} (4p);
    \draw[latex'-latex'] (8p) -- node [above right, pos=0.35] {$1$} (4p);

\end{tikzpicture}
} }
\caption{Instability of betweenness centrality $C_B$. The distance between $G$ and $H$ vanishes with decreasing $\epsilon$, however $C^G_B(x_1)=9$ and $C^{H}_B(x_1)=0$ for every $\epsilon>0$.}
\label{fig_centrality_instabiliy_example}
\end{figure*}
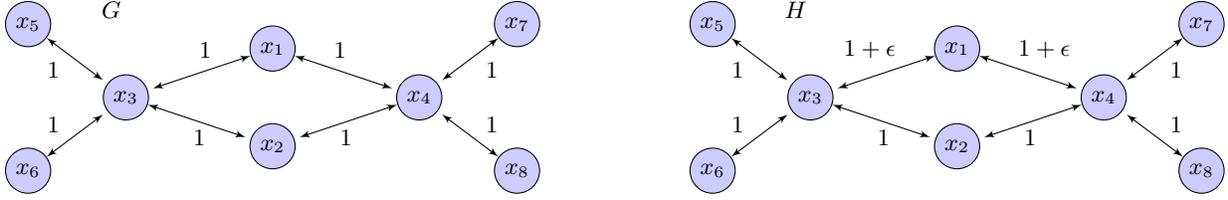
%
%%%%%%%%%%%%%%%%%%%%%%%%%%%%%%%%%%%%%%%%%%%%%%%%%%%%%%%%%%%%%%%%%%%

%%%%%%%%%%%%%%%%%%%%%%%   P   R   O   P   O   S   I   T   I   O   N   %%%%%%%%%%%%%%%%%%%%%%%%%%%%
%
\begin{proposition}\label{prop_centrality_unstable}
The betweenness centrality measure $C_B$ in \eqref{eqn_betweenness_centrality} is not stable in the sense of Definition \ref{def_stability}.
\end{proposition}
%
%%%%%%%%%%%%%%%%%%%%%%%%%%%%   P   R   O   O   F    %%%%%%%%%%%%%%%%%%%%%%%%%%%%%%%
\begin{myproofnoname}
Consider the undirected graphs $G=(V, E, W)$ and $H=(V, E, W')$ depicted in Fig. \ref{fig_centrality_instabiliy_example}. Since the sum in \eqref{eqn_metric_definition} is done over the set of directed edges, it is immediate that $d_{(V,E)}(G,H)= 4 \epsilon$.

For any $\epsilon > 0$, according to \eqref{eqn_betweenness_centrality} we have that $C^G_B(x_1)=9$ since the node $x_1$ is part of one of the two shortest paths from any node in $\{x_3, x_5, x_6\}$ to any node in $\{x_4, x_7, x_8\}$ and vice versa, where the other path goes through $x_2$. However, for that same $\epsilon$, $C^{H}_B(x_1)=0$ since $x_1$ is not an intermediate node in any shortest path in graph $H$. This implies that,
\begin{equation}\label{eqn_instability_example}
\frac{|C^G_B(x_1)-C^{H}_B(x_1)|}{d_{(V,E)}(G,H)} = \frac{9}{4 \epsilon}.
\end{equation}
Note that constant $K_G$ in \eqref{eqn_stability_definition} cannot depend on $\epsilon$ since this is not a parameter of the graph $G$. Thus, for any candidate constant $K_G$ there exists a small enough $\epsilon > 0$ such that the above ratio is greater than the proposed $K_G$. Thus, such constant cannot exist and $C_B$ is not stable.
\end{myproofnoname}
%%%%%%%%%%%%%%%%%%%%%%%%%%%%%%%%%%%%%%%%%%%%%%%%%%%%%%%%%%%%%%%%%%%%

The instability of the betweenness centrality measure entails an undesirable behavior when applied to synthetic and real-world networks as shown in Section \ref{sec_numerical_experiments}. Also, the result in Proposition \ref{prop_centrality_unstable} motivates an alternative definition of betweenness centrality presented in Section \ref{sec_stable_centrality_measure}.

\subsection{Eigenvector centrality}\label{sec_eigenvector_centrality}

The eigenvector centrality $C_E$ of a node, just as the degree centrality, depends on its neighbors. However, it does not depend on the number of neighbors but rather on how important its neighbors are. The importance of its neighbors in turn depends on how important \emph{their} neighbors are, and so on. In this way, a node with a few important neighbors has larger eigenvector centrality than a node with various neighbors of limited importance. Following this premise, for a given graph $G=(V,E,W)$ with adjacency matrix $A$ where weights denote similarities, we may write for every node $x$
\begin{equation}\label{eqn_towards_eigen_centrality_010}
C_E(x) := \frac{1}{\lambda} \sum_{(x, x') \in E} W(x, x') C_E(x'),
\end{equation}
for some constant $\lambda$. In \eqref{eqn_towards_eigen_centrality_010}, the centrality value of a node is defined as a weighted average of the centrality values of its neighbors. In terms of the adjacency matrix, we have that
\begin{equation}\label{eqn_towards_eigen_centrality_020}
C_E(x_i) = \frac{1}{\lambda} \sum_{j} A_{ij} C_E(x_j).
\end{equation}
We may rewrite \eqref{eqn_towards_eigen_centrality_020} in matrix form to obtain \cite{Bonacich72}
\begin{equation}\label{eqn_towards_eigen_centrality_030}
\lambda \, C_E = A \, C_E,
\end{equation}
where $C_E = (C_E(x_1), \ldots, C_E(x_n))^T$. From \eqref{eqn_towards_eigen_centrality_030} it is immediate that $C_E$ is an eigenvector of the adjacency matrix $A$. In order to ensure that the components of $C_E$ are real numbers, $A$ must be symmetric which corresponds to graph $G$ being undirected. Although some extensions have been proposed for directed graphs \cite{Bonacich01}, the most commonly used version of eigenvector centrality requires the graph to be undirected. The solution of \eqref{eqn_towards_eigen_centrality_030} is not uniquely determined, since every pair $(\lambda, C_E)$ of eigenvalues and eigenvectors solves the equation. However, for connected graphs the Perron-Frobenius Theorem ensures that the eigenvector corresponding to the maximal eigenvalue contains all positive components; see Lemma \ref{lem_perron_frobenius} below. Thus, $C_E$ in \eqref{eqn_towards_eigen_centrality_030} is defined as the normalized dominant eigenvector of $A$ where the corresponding graph $G$ must be connected and undirected. As a consequence, $C_E(x)$ is upper bounded by 1 for every node $x$. Eigenvector centrality is a stable measure as the following proposition shows.

%%%%%%%%%%%%%%%%%%%%%%%   P   R   O   P   O   S   I   T   I   O   N   %%%%%%%%%%%%%%%%%%%%%%%%%%%%
%
\begin{proposition}\label{prop_eigenvector_centrality_stable}
The eigenvector centrality measure $C_E$ in \eqref{eqn_towards_eigen_centrality_030} is stable as defined in Definition \ref{def_stability} with 
\begin{equation}\label{eqn_k_g_eigenvector_centrality}
K_G = \frac{4}{\lambda_n - \lambda_{n-1}},
\end{equation}
where $\lambda_n > \lambda_{n-1} \geq \ldots \geq \lambda_1$ are the eigenvalues of the adjacency matrix of graph $G$.
\end{proposition}
In proving Proposition \ref{prop_eigenvector_centrality_stable}, we will use as lemmas two known results from linear algebra. The first result is the Perron-Frobenius Theorem which we restate below in a form that is useful for our proof.

%%%%%%%%%%%%%%%%%%%%%%%%%%   L   E   M   M   A   %%%%%%%%%%%%%%%%%%%%%%%%%%%%%%%%%
\begin{lemma}\label{lem_perron_frobenius}
{\bf [Perron-Frobenius Theorem]} Let $A \geq 0$ be irreducible. Then there is a unique positive real number $r$ such that:
\begin{itemize}
\item[(i)] There is a real vector $\bbv > 0$ with $A\bbv = r \bbv$.
\item[(ii)] The geometric and algebraic multiplicities of $r$ are one.
\item[(iii)] For each eigenvalue $s$ of $A$ we have that $|s| \leq r$.
\end{itemize}
\end{lemma}
%%%%%%%%%%%%%%%%%%%%%%%%%%%%%%%%%%%%%%%%%%%%%%%%%%%%%%%%%%%%%%%%%%%%
\begin{myproofnoname}
See \cite[Chapter 2, Theo. 1.4]{BermanPlemmons94}.
\end{myproofnoname}
%%%%%%%%%%%%%%%%%%%%%%%%%%%%%%%%%%%%%%%%%%%%%%%%%%%%%%%%%%%%%%%%%%%%

The second result studies the behavior of eigenvectors when a symmetric matrix is perturbed.

%%%%%%%%%%%%%%%%%%%%%%%%%%   L   E   M   M   A   %%%%%%%%%%%%%%%%%%%%%%%%%%%%%%%%%
\begin{lemma}\label{lem_eigenvector_perturbation}
{\bf [Eigenvector Perturbation Theorem]} Let $A$ and $A+E$ in $\reals ^{n \times n}$ be symmetric with eigenvalues $\lambda_n \geq \ldots \geq \lambda_1$ and $\mu_n \geq \ldots \geq \mu_1$. If
\begin{equation}\label{eqn_lem_eigenvector_010}
|\lambda_i - \lambda_j| \geq \beta > ||E||_2, \quad i \neq j, 
\end{equation} 
then $A$ and $A+E$ have normalized eigenvectors $\bbu_j$ and $\bbv_j$ corresponding to $\lambda_j$ and $\mu_j$ such that
\begin{equation}\label{eqn_lem_eigenvector_020}
||\bbu_j - \bbv_j ||_2 \leq \gamma (1+ \gamma^2)^{1/2},
\end{equation} 
where $\gamma = ||E||_2/(\beta - ||E||_2)$.
\end{lemma}
%%%%%%%%%%%%%%%%%%%%%%%%%%%%%%%%%%%%%%%%%%%%%%%%%%%%%%%%%%%%%%%%%%%%
\begin{myproofnoname}
See \cite[Theorem 3.3.7]{Ortega90}.
\end{myproofnoname}
%%%%%%%%%%%%%%%%%%%%%%%%%%%%%%%%%%%%%%%%%%%%%%%%%%%%%%%%%%%%%%%%%%%%

We now use the results in Lemmas \ref{lem_perron_frobenius} and \ref{lem_eigenvector_perturbation} to show the stability of the eigenvector centrality measure.

%%%%%%%%%%%%%%%%%%   P   R   O   O   F      O   F      P   R   O   P   O   S   I   T   I   O   N    %%%%%%%%%%%%%%%%%%%%
\begin{myproof}[of Proposition \ref{prop_eigenvector_centrality_stable}]
Consider an arbitrary undirected, connected graph $G = (V, E, W)$ with adjacency matrix $A$ and another undirected graph $H=(V,E,W')$ with the same set of edges and adjacency matrix $B$. The following result relating the distance between $G$ and $H$ and the norm of the difference of adjacency matrices will be useful for the rest of the proof.
%%%%%%%%%%%%%%%%%%%%%%%%%   C   L   A   I   M   %%%%%%%%%%%%%%%%%%%%%%%%%%%%%%%%%%%
\begin{claim}
The distance between the graphs upper bounds the $2-$norm of the difference between the adjacency matrices, i.e.
\begin{equation}\label{eqn_claim_distance_bounds_norm}
d_{(V,E)}(G, H) \geq ||B-A||_2.
\end{equation}
\end{claim}
\begin{myproof}
From the definition of $d_{(V,E)}$ in terms of adjacency matrices \eqref{eqn_metric_definition}, we may write that
\begin{align}\label{eqn_claim_distance_bounds_010}
d_{(V,E)}(G, H) & = \sum_{i,j} |B_{ij} - A_{ij}| \geq \sqrt{\sum_{i,j} (B_{ij} - A_{ij})^2} \nonumber \\ 
& =  \sqrt{\text{Trace}\left( (B-A)^T(B-A) \right)},
\end{align}
where the inequality is given by the relation between the $\ell_1$ and $\ell_2$ vector norms. Noticing that $(B-A)^T(B-A)$ is a positive semi-definite matrix and the fact that the trace equals the sum of the eigenvalues, it follows that
\begin{align}\label{eqn_claim_distance_bounds_020}
d_{(V,E)}(G, H) \geq \sqrt{\lambda_{\max}\left( (B-A)^T(B-A) \right)}.
\end{align}
Observing that the right hand side of \eqref{eqn_claim_distance_bounds_020} is exactly $||B-A||_2$ concludes the proof.
\end{myproof}
%%%%%%%%%%%%%%%%%%%%%%%%%%%%%%%%%%%%%%%%%%%%%%%%%%%%%%%%%%%%%%%%%%%%

Continuing with the main proof of Proposition \ref{prop_eigenvector_centrality_stable}, denote by $\lambda_n \geq \ldots \geq \lambda_1$ the eigenvalues of $A$ with corresponding eigenvectors $\bbv_n, \ldots, \bbv_1$ and, similarly for $B$ where the eigenvalues are $\mu_i$ and the associated eigenvector $\bbu_i$, for $i=1, \ldots, n$. From the connectedness of graphs $G$ and $H$, it follows that matrices $A$ and $B$ are irreducible (cf. \cite[Chapter 2, Theo. 2.7]{BermanPlemmons94}). Thus, by Lemma \ref{lem_perron_frobenius}, the dominant eigenvalues of both matrices $\lambda_n$ and $\mu_n$ must be simple. Denote by $\delta$ the distance from the dominant eigenvalue in $A$ to the second largest eigenvalue, i.e. $\delta := \lambda_n - \lambda_{n-1}$. For simplicity, we divide the proof of Proposition \ref{prop_eigenvector_centrality_stable} into two cases. In the first case, we assume that 
\begin{equation}\label{eqn_proof_eigencentrality_010}
||B-A||_2 \geq \frac{\delta}{4}.
\end{equation}
For this case, pick an arbitrary node $x \in V$ and, given that the eigenvector centrality is bounded between 0 and 1 we can write that
\begin{align}\label{eqn_proof_eigencentrality_015}
|C^G_E(x)-C^H_E(x)| & \leq 1  =\frac{4}{\delta} \,\, \frac{\delta}{4} \leq \frac{4}{\delta} \,\, ||B-A||_2 \nonumber\\
& \leq \frac{4}{\delta} \,\, d_{(V,E)}(G, H),
\end{align}
where we used \eqref{eqn_proof_eigencentrality_010} and \eqref{eqn_claim_distance_bounds_norm}. This shows inequality \eqref{eqn_stability_definition} for $K_G=4/\delta$ in the first case studied.

Consider as a second scenario the situation where 
\begin{equation}\label{eqn_proof_eigencentrality_020}
||B-A||_2 < \frac{\delta}{4}.
\end{equation}
Notice that this implies that \eqref{eqn_lem_eigenvector_010} is satisfied for $j=n$, $\beta = \delta$ and $E = B - A$. For any given node $x \in V$, we have that 
\begin{equation}\label{eqn_proof_eigencentrality_030}
|C^G_E(x) - C^H_E(x)| = | \bbv_n(x) - \bbu_n(x) | \leq || \bbv_n - \bbu_n ||_2,
\end{equation}
where the equality comes from the definition of eigenvector centrality [cf. \eqref{eqn_towards_eigen_centrality_030}] and the inequality is a trivial fact from vector algebra. Combining the result of Lemma \ref{lem_eigenvector_perturbation} with \eqref{eqn_proof_eigencentrality_030}, we can write
\begin{align}\label{eqn_proof_eigencentrality_040}
|C^G_E&(x) - C^H_E(x)| \\ 
& \leq \frac{||B-A||_2}{(\delta - ||B-A||_2)} \left( 1 + \frac{||B-A||^2_2}{(\delta - ||B-A||_2)^2} \right)^{1/2} \nonumber \\
& =  \frac{(\delta^2 - 2 \delta \, ||B-A||_2 + 2 \, ||B-A||^2_2)^{1/2}}{(\delta - ||B-A||_2)^2} \,\,\, ||B-A||_2.  \nonumber
\end{align}
Using the assumed relation in \eqref{eqn_proof_eigencentrality_020}, the numerator in the right hand side of \eqref{eqn_proof_eigencentrality_040} can be upper bounded by $\delta$ and the denominator can be lower bounded by $(\delta/2)^2$. Thus, it follows that
\begin{align}\label{eqn_proof_eigencentrality_050}
|C^G_E(x)& - C^H_E(x)| \leq \frac{4 \delta}{\delta^2} \,\, ||B-A||_2 \leq \frac{4}{\delta} \,\,\, d_{(V,E)}(G, H),
\end{align}
where we used \eqref{eqn_claim_distance_bounds_norm} in the last inequality. This shows inequality \eqref{eqn_stability_definition} for $K_G=4/\delta$ in the second case analyzed. Since for every pair of graphs $G$ and $H$, either \eqref{eqn_proof_eigencentrality_010} or \eqref{eqn_proof_eigencentrality_020} must be satisfied, inequality \eqref{eqn_stability_definition} for $K_G=4/\delta$ is true in general, showing that the eigenvector centrality measure is stable.
\end{myproof}
%%%%%%%%%%%%%%%%%%%%%%%%%%%%%%%%%%%%%%%%%%%%%%%%%%%%%%%%%%%%%%%%%%%%

Notice that the constants $K_G$ found for degree centrality [cf. Proposition \ref{prop_degree_is_stable}] and closeness centrality [cf. Proposition \ref{prop_closeness_is_stable}] only depend on the number of nodes $n$ and are independent of the weight structure of the graph $G$. However, this is not the case for eigenvector centrality, where the constant $K_G$ depends on the eigenvalues of the adjacency matrix which are a function of the weights of the graph. This difference does not impact the practical implementation of eigenvector centrality as we see in Section \ref{sec_numerical_experiments}.

Among the four centrality measures studied -- degree, closeness, betweenness, and eigenvector --, betweenness centrality is the only measure that fails to be stable. This motivates the alternative definition for a stable betweenness centrality that we develop in the following section. 

%%%%%%%%%%%%%%%%%%%%%%%%%%%%%%%%%%%%%%%%%%%%%%%%%%%%%%%%%%%%%%%%%%%%%

%%%%%%%%%%%%%%%%%%%%%%%%%%%%%%%%%%%%%%%%%%%%%%%%%%%%%%%%%%%%%%%%%%%
%%%%%%   S   T   A   B   I   L  I   Z    I    N    G         B    E    T    W   E    E    N   N   E   S    S        C   E   N    T    R   A   L   I   T   Y   %%%%%%%
%%%%%%%%%%%%%%%%%%%%%%%%%%%%%%%%%%%%%%%%%%%%%%%%%%%%%%%%%%%%%%%%%%%
\section{Stable Betweenness Centrality}
\label{sec_stable_centrality_measure}

A consequence of the instability of betweenness centrality shown in Proposition \ref{prop_centrality_unstable} is that perturbations in the weights  of a graph have major impacts on the centrality ranking of the nodes; see Section \ref{sec_numerical_experiments}. Thus, in this section we present an alternative centrality measure that preserves the centrality notion of betweenness centrality while being stable.

Given an arbitrary graph $G=(V,E,W)$ and a node $x \in V$, define a new graph $G^x=(V^x, E^x, W^x)$ with $V^x=V \backslash \{x\}$, $E^x = E \, \backslash \{ (x', x'') \, | \, x'=x \,\, \text{or} \,\,  x''=x \}$, and $W^x = W|_{E^x}$. I.e., the graph $G^x$ is constructed by deleting from $G$ the node $x$ and every edge directed to or from it. Define the stable betweenness centrality $C_{SB}(x)$ of any node $x \in V$ as
\begin{equation}\label{eqn_stable_centrality_definition}
C_{SB}(x) := \sum_{\substack{x', x'' \in V \\ x' \neq x \neq x'' }} s_{G^x}(x', x'')-s_{G}(x', x'').
\end{equation}
Note that every term in the above summation is nonnegative since shortest paths in the graph $G^x$ cannot be shorter than the corresponding paths in $G$. Measure $C_{SB}$ quantifies the centrality of a given node $x$ by the change in the length of shortest paths once this node is removed. Intuitively, if a node is part of many shortest paths, when we remove this node the corresponding paths will increase in length and result in a high centrality value. In this sense, measure $C_{SB}$ is similar to the original betweenness centrality measure $C_B$. However, how critical a given node is in connecting the network depends on the best alternative path if this node fails. As was the case for $C_B$, definition \eqref{eqn_stable_centrality_definition} should be applied to graphs where the weights represent dissimilarities between nodes. In contrast to the traditional centrality measure, $C_{SB}$ is stable as shown after the following remark.

\begin{remark}\label{rem_stable_betweenness}\normalfont
To guarantee that $C_{SB}$ in \eqref{eqn_stable_centrality_definition} will achieve finite values, we must require that the graph being studied is 2-connected or biconnected \cite{West01}. In this way, the shortest path length $s_{G^x}(x', x'')$ is finite for all triplets $x, x', x'' \in V$. An alternative is to adopt the convention that $\infty - \infty = 0$ and in such a case no assumption needs to be made about the connectivity of the graph [cf. Remark \ref{rem_closeness_centrality}].
\end{remark}

%
%%%%%%%%%%%%%%%%%%%%%%%   P   R   O   P   O   S   I   T   I   O   N   %%%%%%%%%%%%%%%%%%%%%%%%%%%%
%
\begin{proposition}\label{prop_stable_centrality_stable}
The stable betweenness centrality measure $C_{SB}$ in \eqref{eqn_stable_centrality_definition} is stable as defined in Definition \ref{def_stability} with $K_G= 2 \,  n^2$.
\end{proposition}
In proving this proposition, we use the following lemma.
%
%%%%%%%%%%%%%%%%%%%%%%%%%%    L   E   M   M   A   %%%%%%%%%%%%%%%%%%%%%%%%%%%%%%%%%
\begin{lemma}\label{lem_stable_centrality_stable}
Given two arbitrary graphs $G=(V,E,W)$ and $H=(V,E,W')$ we have that 
\begin{equation}\label{eqn_lem_stable_centrality_stable}
d_{(V,E)}(G, H) \geq d_{(V^x,E^x)}(G^x, H^x),
\end{equation}
for all $x \in V$.
\end{lemma}
%%%%%%%%%%%%%%%%%%%%%%%%%%%    P R O O F   %%%%%%%%%%%%%%%%%%%%%%%%%%%%%%%%%%
\begin{myproofnoname}
Use the definition of $d_{(V,E)}$ in \eqref{eqn_metric_definition} and separate all the terms that involve $x$ to write
\begin{align}\label{eqn_proof_lem_stable_centrality_stable_010}
 d_{(V,E)}(G, H) & = \sum_{e \in E} |W(e)-W'(e)| \\ \nonumber 
 &= \sum_{\substack{(x', x'') \in E \\ x' \neq x \neq x'' }}  |W(x',x'') - W'(x',x'')| \\ \nonumber &\qquad + 
\sum_{\substack{(x', x'') \in E \\ x = x'  \, \text{or}\,  x = x'' }} |W(x',x'')-W'(x',x'')|
\end{align}
The first term in the rightmost side of the equality in \eqref{eqn_proof_lem_stable_centrality_stable_010} is, by definition, the distance $d_{(V^x,E^x)}(G^x, H^x)$. We can then rewrite \eqref{eqn_proof_lem_stable_centrality_stable_010} as
\begin{align}\label{eqn_proof_lem_stable_centrality_stable_020}
   d_{(V,E)}(G, H) & = d_{(V^x,E^x)}(G^x, H^x) &  \\ \nonumber & \qquad +
        \sum_{\substack{(x', x'') \in E \\ x = x'  \, \text{or}\,  x = x'' }} \! |W(x',x'')-W'(x',x'')|,
\end{align}
The result in \eqref{eqn_lem_stable_centrality_stable} follows because the second term in \eqref{eqn_proof_lem_stable_centrality_stable_020} is nonnegative.
\end{myproofnoname}

We now use Lemmas \ref{lem_shortest_path_less_distance} and \ref{lem_stable_centrality_stable} to prove Proposition \ref{prop_stable_centrality_stable}.

%%%%%%%%%%%%%%%%%%%%%%  P R O O F    O F    P R O P O S I T I O N  %%%%%%%%%%%%%%%%%%%%%%%%% 
\begin{myproof}[of Proposition \ref{prop_stable_centrality_stable}]
Given two biconnected graphs $G = (V, E, W)$ and $H = (V, E, W')$ we have that for an arbitrary node $x \in V$,
\begin{align}\label{eqn_proof_stable_centrality_stable_010}
|C_{SB}^G(x) & - C_{SB}^H(x)|  = \Big| \sum_{\substack{x', x'' \in V \\ x' \neq x \neq x'' }} s_{G^x}(x', x'')-s_{G}(x', x'') - \nonumber \\
& \sum_{\substack{x', x'' \in V \\ x' \neq x \neq x'' }} s_{H^x}(x', x'')-s_{H}(x', x'') \Big|.
\end{align}
Rearranging terms and using the triangle inequality we obtain
\begin{align}\label{eqn_proof_stable_centrality_stable_020}
&|C_{SB}^G(x) - C_{SB}^H(x)|  \leq  \\
&  \sum_{\substack{x', x'' \in V \\ x' \neq x \neq x'' }} \!\!\! \left| s_H(x', x'')-s_{G}(x', x'') \right| \!\! + \!\! \left| s_{G^x}(x', x'')-s_{H^x}(x', x'') \right|\!\! \nonumber
\end{align}
Applying Lemma \ref{lem_shortest_path_less_distance} to \eqref{eqn_proof_stable_centrality_stable_020} we have that
\begin{align}\label{eqn_proof_stable_centrality_stable_030}
|C_{SB}^G(x) - &C_{SB}^H(x)| \\ \nonumber 
&\leq  \sum_{\substack{x', x'' \in V \\ x' \neq x \neq x'' }} d_{(V,E)}(G, H) + d_{(V^x,E^x)}(G^x, H^x) \\ \nonumber
& \leq  n^2 \, \left(d_{(V,E)}(G, H) + d_{(V^x,E^x)}(G^x, H^x)\right).
\end{align}
Using now Lemma \ref{lem_stable_centrality_stable} we obtain that
\begin{align}\label{eqn_proof_weighted_centrality_stable_080}
|C_{SB}^G(x) - C_{SB}^H(x)| \leq 2 n^2 \, d_{(V,E)}(G, H),
\end{align}
showing inequality \eqref{eqn_stability_definition} for $K_G=2 n^2$ and concluding the proof.
\end{myproof}
%%%%%%%%%%%%%%%%%%%%%%%%%%%%%%%%%%%%%%%%%%%%%%%%%%%%%%%%%%%%%%%%%%%

\begin{figure*}
\centering
\centerline{\def \thisplotscale {0.65}
\def \unit {\thisplotscale cm}

{\small
\begin{tikzpicture}[-stealth, shorten >=2, scale = \thisplotscale]

%%% FIRST NETWORK %%%%%%%%%
% NAME
    \node at (0-3-3.3,0.8) {$G$};
    
% NODES

    \node [blue vertex, minimum height=0.6cm, minimum width=0.6cm] at (0-3,0) (1) {$x_1$};
    \node [blue vertex, minimum height=0.6cm, minimum width=0.6cm] at (0-3,-2) (2) {$x_2$};    
    \node [blue vertex, minimum height=0.6cm, minimum width=0.6cm] at (-3-3,-1) (3) {$x_3$};
    \node [blue vertex, minimum height=0.6cm, minimum width=0.6cm] at (3-3,-1) (4) {$x_4$};
    \node [blue vertex, minimum height=0.6cm, minimum width=0.6cm] at (-4-4,0.5) (5) {$x_5$};
    \node [blue vertex, minimum height=0.6cm, minimum width=0.6cm] at (-4-4,-2.5) (6) {$x_6$};
    \node [blue vertex, minimum height=0.6cm, minimum width=0.6cm] at (4-2,0.5) (7) {$x_7$};
    \node [blue vertex, minimum height=0.6cm, minimum width=0.6cm] at (4-2,-2.5) (8) {$x_8$};

%EDGES
    
    \draw[latex'-latex'] (1) -- node [pos=0.3, above left] {$1$} (3);
    \draw[latex'-latex'] (1) -- node [pos=0.3, above right] {$1$} (4);
    \draw[latex'-latex'] (3) -- node [below left, pos=0.7] {$1+ \epsilon$} (2);
    \draw[latex'-latex'] (4) -- node [below right, pos=0.7] {$1 + \epsilon$} (2);
    \draw[latex'-latex'] (5) -- node [below left, pos=0.35] {$1$} (3);
    \draw[latex'-latex'] (6) -- node [above left, pos=0.35] {$1$} (3);
    \draw[latex'-latex'] (7) -- node [below right, pos=0.35] {$1$} (4);
    \draw[latex'-latex'] (8) -- node [above right, pos=0.35] {$1$} (4);

    %%% SECOND NETWORK %%%%%%%%%
% NAME
    \node at (7.7,0.8) {$H$};
    
% NODES

    \node [blue vertex, minimum height=0.6cm, minimum width=0.6cm] at (11,0) (1p) {$x_1$};
    \node [blue vertex, minimum height=0.6cm, minimum width=0.6cm] at (11,-2) (2p) {$x_2$};    
    \node [blue vertex, minimum height=0.6cm, minimum width=0.6cm] at (8,-1) (3p) {$x_3$};
    \node [blue vertex, minimum height=0.6cm, minimum width=0.6cm] at (14,-1) (4p) {$x_4$};
    \node [blue vertex, minimum height=0.6cm, minimum width=0.6cm] at (7-1,0.5) (5p) {$x_5$};
    \node [blue vertex, minimum height=0.6cm, minimum width=0.6cm] at (7-1,-2.5) (6p) {$x_6$};
    \node [blue vertex, minimum height=0.6cm, minimum width=0.6cm] at (15+1,0.5) (7p) {$x_7$};
    \node [blue vertex, minimum height=0.6cm, minimum width=0.6cm] at (15+1,-2.5) (8p) {$x_8$};

%EDGES
    
    \draw[latex'-latex'] (1p) -- node [pos=0.3, above left] {$1$} (3p);
    \draw[latex'-latex'] (1p) -- node [pos=0.3, above right] {$1$} (4p);
    \draw[latex'-latex'] (3p) -- node [below left, pos=0.7] {$1 + M$} (2p);
    \draw[latex'-latex'] (4p) -- node [below right, pos=0.7] {$1 + M$} (2p);
    \draw[latex'-latex'] (5p) -- node [below left, pos=0.35] {$1$} (3p);
    \draw[latex'-latex'] (6p) -- node [above left, pos=0.35] {$1$} (3p);
    \draw[latex'-latex'] (7p) -- node [below right, pos=0.35] {$1$} (4p);
    \draw[latex'-latex'] (8p) -- node [above right, pos=0.35] {$1$} (4p);

\end{tikzpicture}
} }
\caption{Implementation example of betweenness $C_B$ and stable betweenness $C_{SB}$ centrality. The betweenness centrality value of $x_1$ is equal for both graphs, $C^G_B(x_1) = C^H_B(x_1)=18$. However, the stable betweenness is different, $C_{SB}^G(x_1)=36 \, \epsilon$ and $C_{SB}^H(x_1)=36 \, M$. The stable betweenness centrality value of $x_1$ depends on the quality of the best alternative path.}
\label{fig_stable_betweenness_example}
\end{figure*}
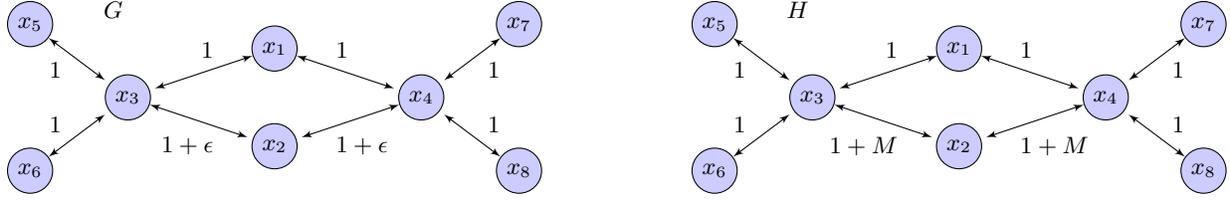

To compare the stable betweenness centrality measure $C_{SB}$ with the traditional measure $C_B$, consider the graphs $G$ and $H$ in Fig. \ref{fig_stable_betweenness_example} where $0 < \epsilon \ll 1 \ll M$, i.e., $\epsilon$ is a small modification to the reference edge weight of $1$ and $M$ is a large modification. For the traditional betweenness centrality we have $C^G_B(x_1) = C^H_B(x_1) = 18$ because $x_1$ is part of 18 shortest paths in both networks [cf. proof of Proposition \ref{prop_centrality_unstable}]. However, intuition suggests that $x_1$ is more central to graph $H$ than it is to graph $G$. A failure of this node in graph $H$ would compromise the graph dynamics deeply since all the flows that passed through $x_1$ are now required to pass through the much costlier edges that run through $x_2$. Graph $G$, however, is more resilient to a failure of $x_1$ because the flows can pass through $x_2$ instead of $x_1$ at similar cost. Thus, it is reasonable to expect $x_1$ to be less central to $G$ than it is to $H$. The stable betweenness centrality $C_{SB}$ captures this notion. If node $x_1$ is deleted from $G$, the 18 shortest paths of which $x_1$ was originally a part of, have their length increased by $2 \epsilon$. Consequently, $C_{SB}^G(x_1)=36 \, \epsilon$. The centrality of node $x_1$ is limited by the existence of a comparable path through node $x_2$. However, if node $x_1$ is deleted from $H$, the 18 shortest paths have their length increased by $2 M$ resulting in $C_{SB}^H(x_1)=36 \, M \gg C_{SB}^G(x_1)$, which corresponds with our intuition. The centrality of $x_1$ depends on the quality of the best alternative.

%%%%%%%%%%%%%%%%%%%%%%%%%%%   F   I   G   U   R   E   %%%%%%%%%%%%%%%%%%%%%%%%%%%%%%%%
%
%%%%%%%%%%%%%%%%%%%%%%%%%%%%%%%%%%%%%%%%%%%%%%%%%%%%%%%%%%%%%%%%%%%

\begin{remark} \normalfont Computing the betweenness centrality $C_B$ for every node in a graph with $n$ nodes and $m$ weighted edges requires $O(nm+n^2 \log n)$ computations \cite{Brandes01}. For $C_{SB}$, we can use the Floyd-Warshall \cite{Floyd62, Warshall62} or the Johnson \cite{Johnson77} algorithm to compute all-pairs shortest paths in a graph. The latter is more suitable for sparse graphs with a computational complexity of $O(nm+n^2 \log n)$. In a naive computation of $C_{SB}$ for every node in the graph, we can compute the shortest paths for all pairs of nodes in the original network and in every network generated when deleting one node at a time. This requires $n+1$ implementations of Johnson algorithm with a total complexity of $O(n^2m+n^3 \log n)$, i.e., a factor of $n$ more than the traditional betweenness centrality. A faster algorithm could exist since, when a node is deleted from the network, only the shortest paths originally containing this node need to be recomputed. The development of this algorithms is beyond the scope of this paper.
\end{remark}

%%%%%%%%%%%%%%%%%%%%%%%%%%%%%%%%%%%%%%%%%%%%%%%%%%%%%%%%%%%%%%%%%%%%%

%%%%%%%%%%%%%%%%%%%%%%%%%%%%%%%%%%%%%%%%%%%%%%%%%%%%%%%%%%%%%%%%%%%
%%%%%%%%%%%%%%%   S   T   A   B   I   L  I   T   Y        A    N    D        C    O    N   T   I   N   U   I   T   Y     %%%%%%%%%%%%%%%
%%%%%%%%%%%%%%%%%%%%%%%%%%%%%%%%%%%%%%%%%%%%%%%%%%%%%%%%%%%%%%%%%%%
\section{Continuity of Centrality Measures}
\label{sec_stability_and_continuity}

Continuity is a subtler notion of how impervious a centrality measure is to noise. Specifically, we define a continuous centrality measure as one in which the centrality values of every node in a given graph are a continuous function of the weights in the edges of this graph as we formally state next.

%%%%%%%%%%%%%%%%%%%%%   D   E   F   I   N   I   T   I   O   N    %%%%%%%%%%%%%%%%%%%%%%%%%%%%%%%
\begin{definition}\label{def_continuity}
Let $G=(V,E,W)$ be an arbitrary graph with adjacency matrix $A$. For every matrix $B$ such that $B_{ij}=0$ if $A_{ij}=0$ and $B + A \geq 0$ element-wise, define the graph $H=(V,E,W')$ whose adjacency matrix is $A+B$. Then, a centrality measure $C$ is continuous if for every $x \in V$,
\begin{equation}\label{eqn_def_continuity}
C^H(x) \rightarrow C^G(x) \quad \text{as} \quad ||B||_2 \rightarrow 0,
\end{equation}
where $C^G(x)$ is the centrality of node $x$ in graph $G$ and similarly for $H$.
\end{definition}
%%%%%%%%%%%%%%%%%%%%%%%%%%%%%%%%%%%%%%%%%%%%%%%%%%%%%%%%%%%%%%%%%%%

In the above definition, matrix $B$ can be interpreted as a perturbation defined on the edges of graph $G$. A continuous centrality measure ensures that as this perturbation vanishes, the centrality values tend to those in graph $G$. Continuity is a weaker notion than stability since the latter implies the former as we show next.

%%%%%%%%%%%%%%%%%%%%%%%%%   P  R  O  P  O  S  I  T  I  O  N   %%%%%%%%%%%%%%%%%%%%%%%%%%%%%
\begin{proposition}\label{prop_stability_implies_continuity}
If a centrality measure $C$ is stable as in Definition \ref{def_stability} then it is continuous as in Definition \ref{def_continuity}.
\end{proposition}
%%%%%%%%%%%%%%%%%%%%%%%%%%   P   R   O   O   F   %%%%%%%%%%%%%%%%%%%%%%%%%%%%%%%%%% 
\begin{myproofnoname}
By the equivalence of matrix norms \cite{HornJohnson12}, it is immediate that as $||B||_2 \rightarrow 0$ then $d_{(V,E)}(G,H) \rightarrow 0$ where $B$, $G$ and $H$ are defined as in Definition \ref{def_continuity}. Thus, if a given measure $C$ is stable, it must satisfy \eqref{eqn_stability_definition} which implies that
\begin{equation}\label{eqn_proof_stability_implies_continuity}
\left| C^G(x) - C^H(x) \right| \rightarrow 0   \quad \text{as} \quad ||B||_2 \rightarrow 0,
\end{equation}
which is equivalent to the definition in \eqref{eqn_def_continuity}, concluding the proof.
\end{myproofnoname}
%%%%%%%%%%%%%%%%%%%%%%%%%%%%%%%%%%%%%%%%%%%%%%%%%%%%%%%%%%%%%%%%%%%%%

As stated in Section \ref{sec_node_centrality_and_stability}, a centrality measure is a function of a graph that assigns a nonnegative real number to each node. This broad definition enables the existence of a wide variety of measures. In particular, there can exist centrality measures which are continuous but not stable, as we show next.

%%%%%%%%%%%%%%%%%%%%%%%%%   P  R  O  P  O  S  I  T  I  O  N   %%%%%%%%%%%%%%%%%%%%%%%%%%%%%
\begin{proposition}\label{prop_continuity_does_not_imply_stability}
If a centrality measure $C$ is continuous as in Definition \ref{def_continuity} then it need not be stable as in Definition \ref{def_stability}.
\end{proposition}
%%%%%%%%%%%%%%%%%%%%%%%%%%   P   R   O   O   F   %%%%%%%%%%%%%%%%%%%%%%%%%%%%%%%%%% 
\begin{myproofnoname}
For an arbitrary graph $G=(V,E,W)$, consider the degree squared centrality measure $C_{DS}$ such that for every node $x \in V$,
\begin{equation}\label{eqn_def_sq_degree}
C^G_{DS}(x) := \sum_{x' |(x, x') \in E} \big( W(x, x') \big)^2.
\end{equation}
In the above measure, for every node we assign a centrality value equal to the sum of the squares of the weights of incident edges instead of just summing the weights as in degree centrality. This is a valid centrality measure which is continuous but not stable. Continuity follows immediately from the fact that $C_{DS}$ is defined as the sum of quadratic -- hence, continuous -- functions of the weights in the graph. Thus, vanishing perturbations of the weights must have vanishing effect on $C_{DS}$.

To see that $C_{DS}$ is not stable, consider two particular undirected graphs $G=(V, E, W)$ and $H=(V,E,W')$ with two nodes, $V=\{x, x'\}$ and weights $W(x, x')=W(x', x)=1$ and $W'(x, x')=W'(x', x)=1+\delta$ for $\delta > 0$. From definition \eqref{eqn_metric_definition} we have that $d_{(V,E)}(G,H)=2\delta$ and from \eqref{eqn_def_sq_degree}, we obtain $C_{DS}^G(x) = 1$ and $C^H_{DS}(x) = (1+\delta)^2$. Thus, for $C_{DS}$ to be stable the following must be fulfilled [cf. \eqref{eqn_stability_definition}],
\begin{align}\label{eqn_sq_degree_not_stable}
| 1 - (1+\delta)^2 | &\leq K_G \,\, 2 \delta \\
\delta^2 + 2 \delta & \leq K_G \,\, 2 \delta \nonumber\\
\frac{\delta}{2} + 1 & \leq K_G  \nonumber
\end{align}
However, $K_G$ is a constant that does not depend on $\delta$ since this is not a parameter of graph $G$. Thus, for any candidate constant $K_G$, there exists a $\delta$ big enough such that \eqref{eqn_sq_degree_not_stable} is violated, showing that $C_{DS}$ is not stable and concluding the proof.
\end{myproofnoname}
%%%%%%%%%%%%%%%%%%%%%%%%%%%%%%%%%%%%%%%%%%%%%%%%%%%%%%%%%%%%%%%%%%%%%

%%%%%%%%%%%%%%%   F   I   G   U   R    E   :   R    A    N    D        N   E   T   W   O   R   K   %%%%%%%%%%%%%%%%%%%%%
\begin{figure*}
\centering

\begin{subfigure}{.33\textwidth}
  \centering
  \includegraphics[width=\textwidth]{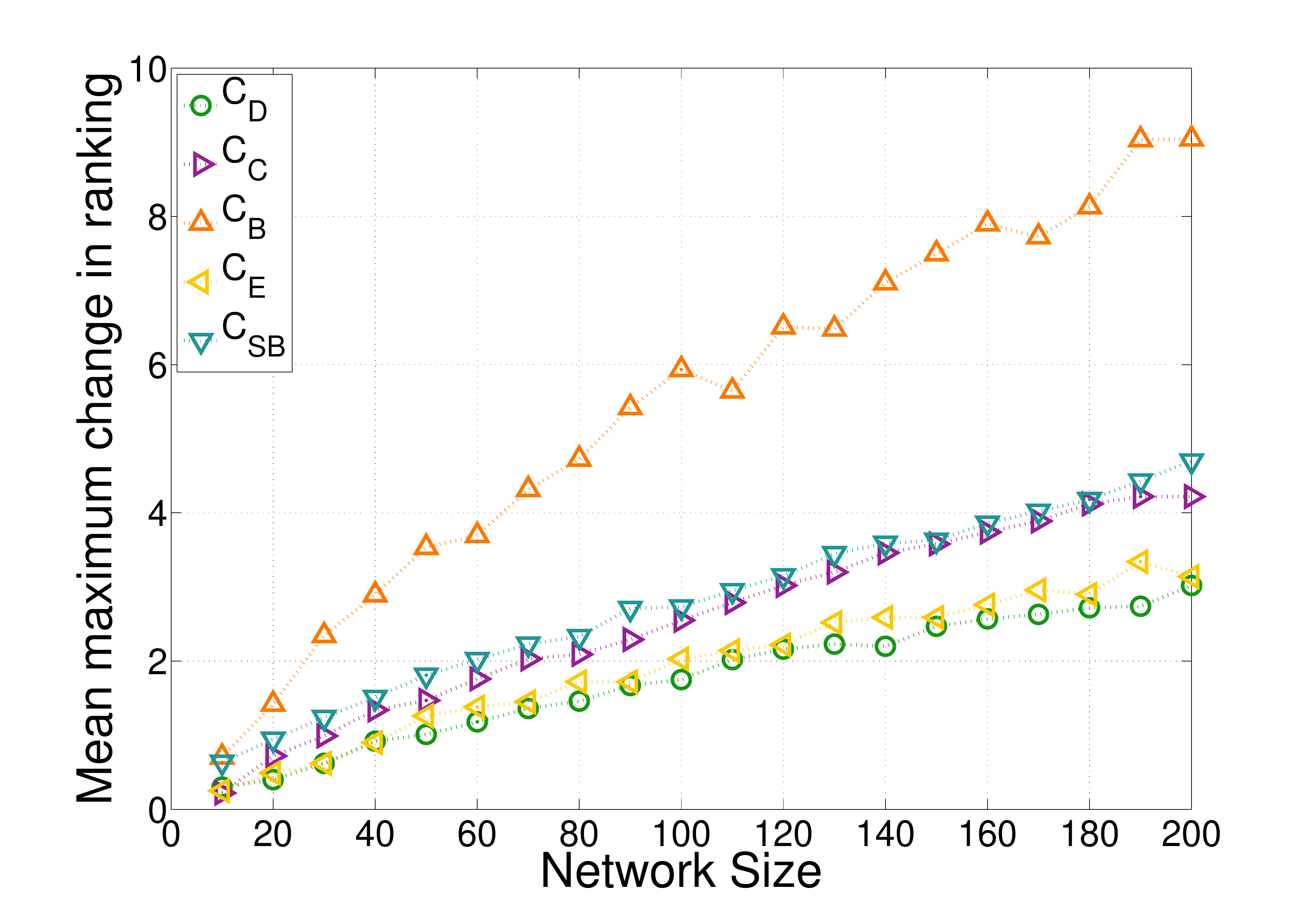}
  \caption{}
  \label{fig:sub1}
\end{subfigure}%
\begin{subfigure}{.33\textwidth}
  \centering
  \includegraphics[width=\textwidth]{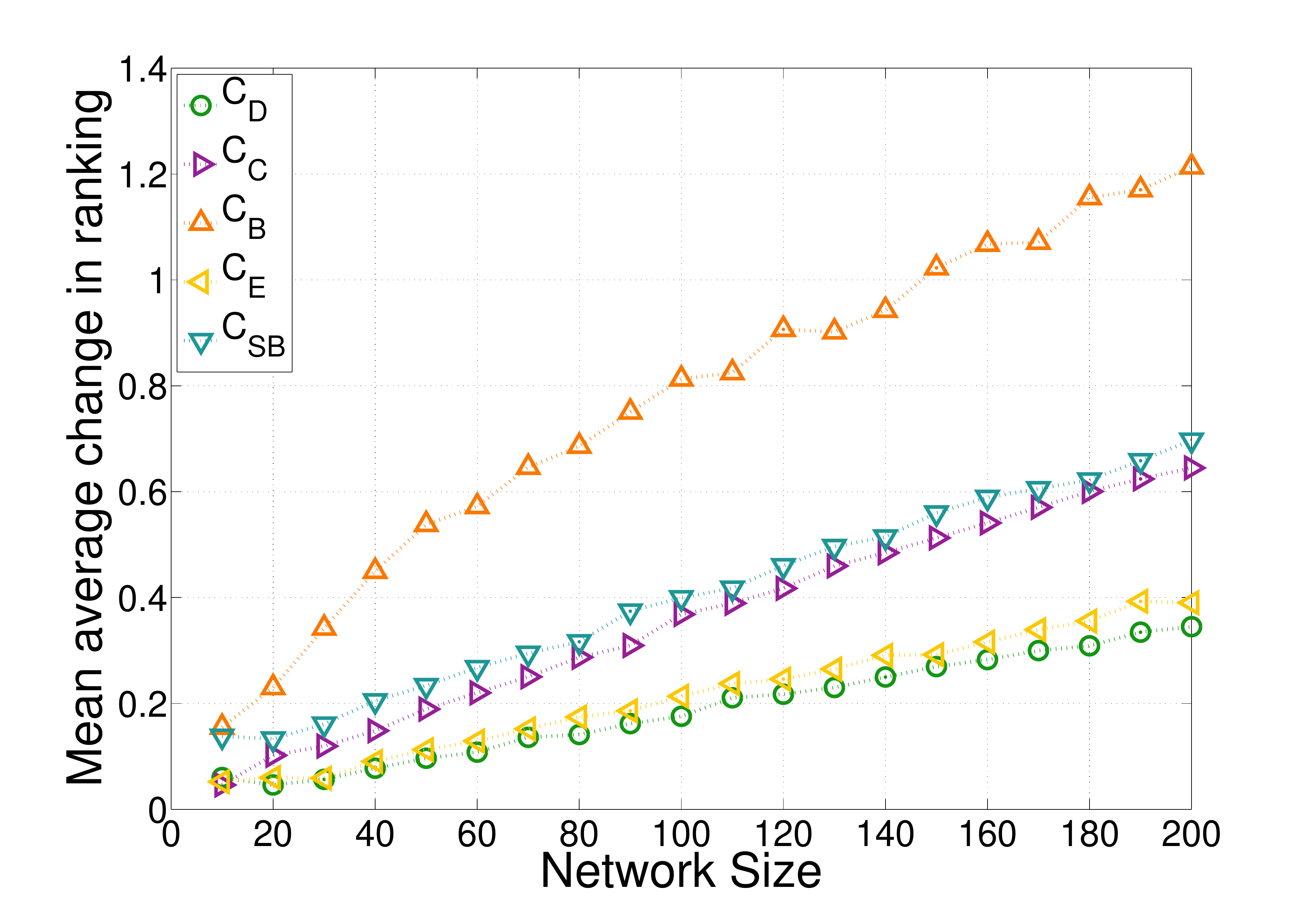}
  \caption{}
  \label{fig:sub2}
\end{subfigure}%
\begin{subfigure}{.33\textwidth}
  \centering
  \includegraphics[width=\textwidth]{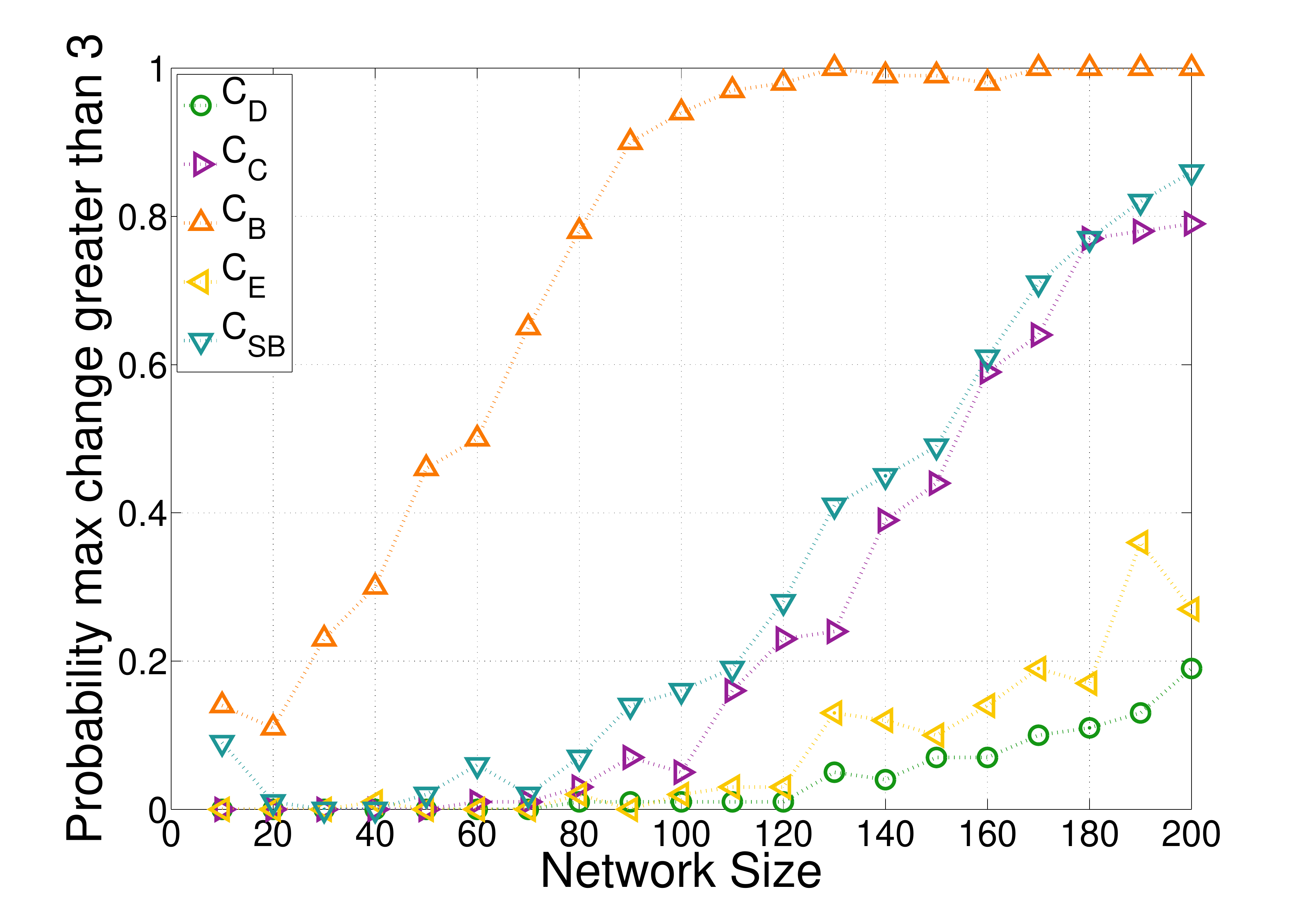}
  \caption{}
  \label{fig:sub3}
\end{subfigure}

\begin{subfigure}{.33\textwidth}
  \centering
  \includegraphics[width=\textwidth]{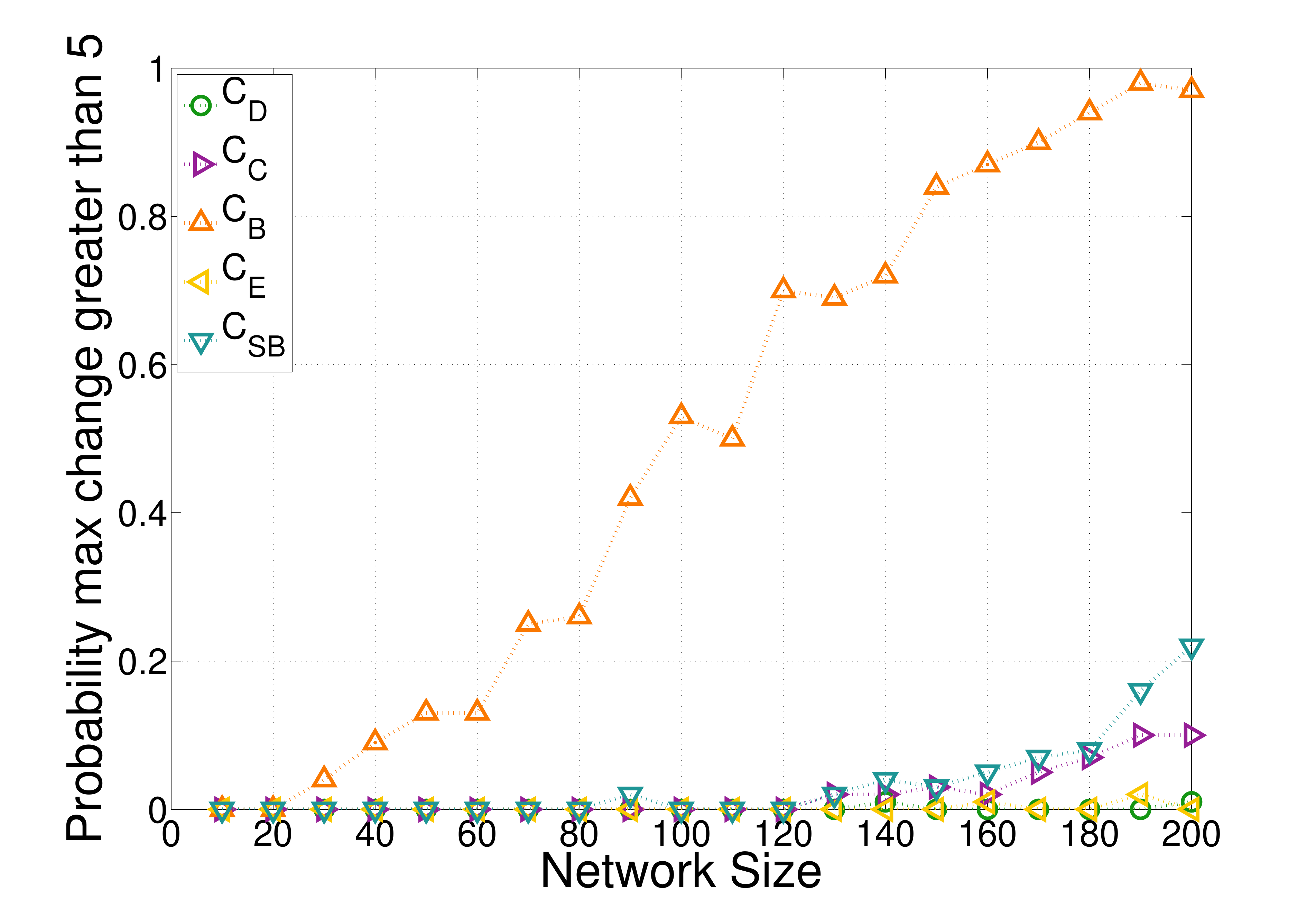}
  \caption{}
  \label{fig:sub4}
\end{subfigure}%
\begin{subfigure}{.33\textwidth}
  \centering
  \includegraphics[width=\textwidth]{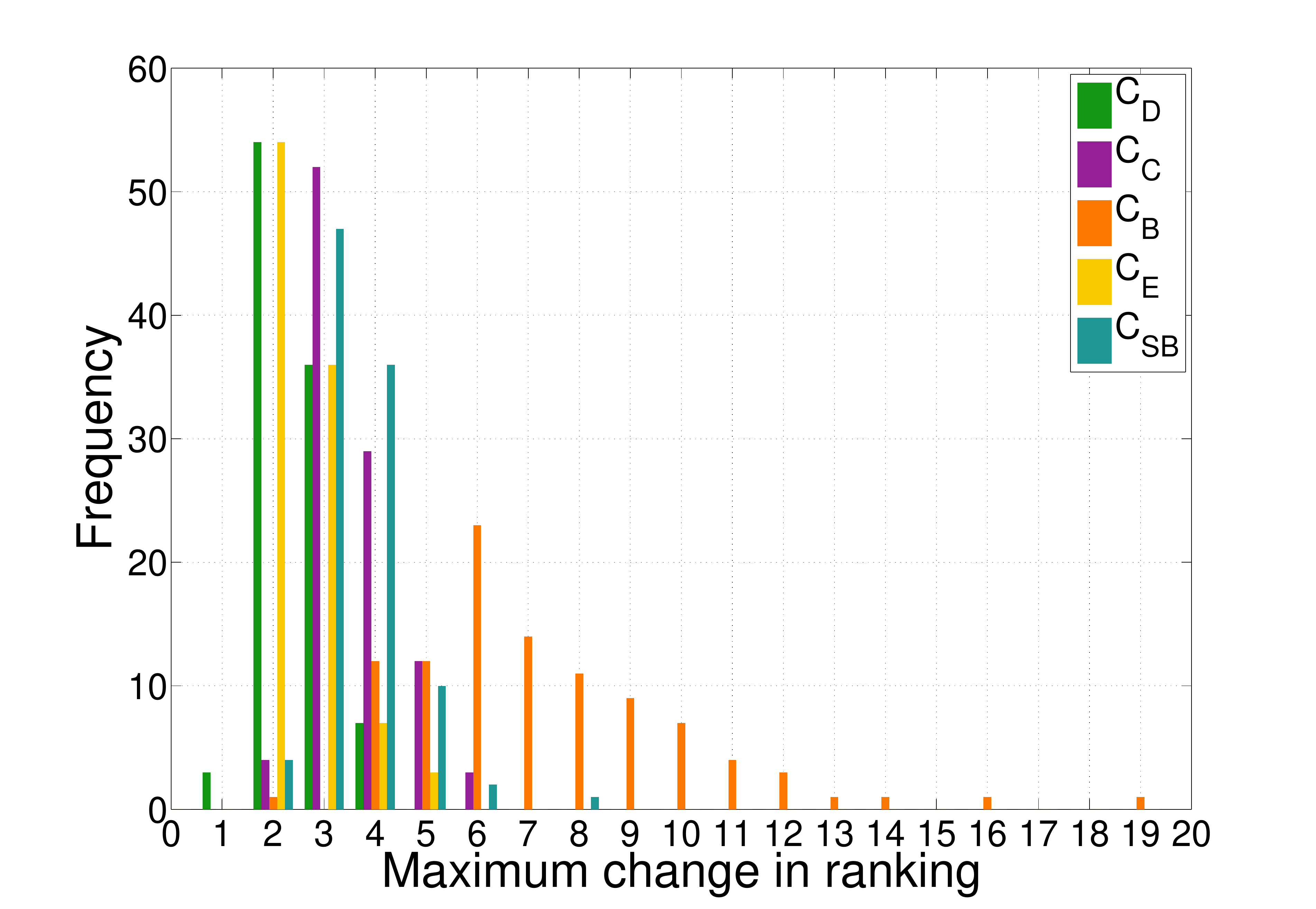}
  \caption{}
  \label{fig:sub5}
\end{subfigure}%
\begin{subfigure}{.33\textwidth}
  \centering
  \includegraphics[width=\textwidth]{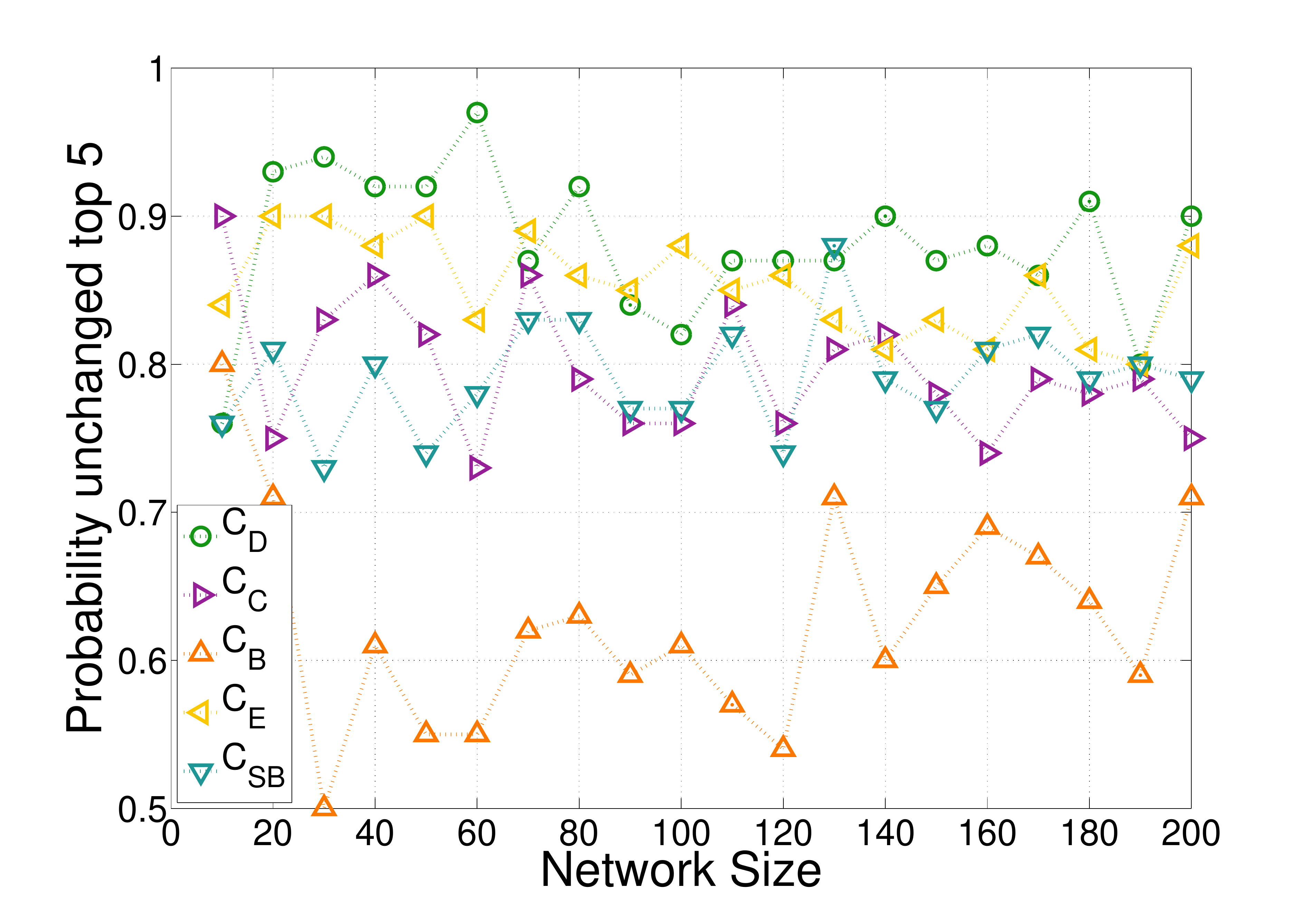}
  \caption{}
  \label{fig:sub6}
\end{subfigure}

\caption{Comparison of stability indicators when type 1 noise ($p_1 = 1$, $\delta_1 = 0.01$) is introduced in random networks for all centrality measures: degree (green circle), closeness (purple right triangle), betweenness (orange upwards triangle),  eigenvector (yellow left triangle), and stable betweenness (cyan downwards triangle). (a) Mean of the maximum change recorded when perturbing a random network as a function of network size. (b) Mean of the average node rank change recorded when perturbing a random network as a function of network size. (c) Probability that the maximum change in the ranking exceeds 3 positions as a function of the network size. (d) Probability that the maximum change in the ranking exceeds 5 positions as a function of the network size. (e) Histogram of the maximum change recorded when perturbing random networks with 150 nodes. (f) Probability that the top 5 ranking remains unchanged when perturbing a network.}
\vspace{-0.1in}
\label{fig_experiments}
\end{figure*}
%%%%%%%%%%%%%%%%%%%%%%%%%%%%%%%%%%%%%%%%%%%%%%%%%%%%%%%%%%%%%%%%%%%

Proposition \ref{prop_stability_implies_continuity} guarantees that degree, closeness, eigenvector and stable betweenness centrality are continuous centrality measures. Proposition \ref{prop_continuity_does_not_imply_stability} leaves open the question of whether betweenness centrality, which is not stable, is continuous or not. The result below shows that it is not.

%%%%%%%%%%%%%%%%%%%%%%%%%   P  R  O  P  O  S  I  T  I  O  N   %%%%%%%%%%%%%%%%%%%%%%%%%%%%%
\begin{proposition}\label{prop_betw_centrality_not_continuous}
The betweenness centrality measure $C_B$ in \eqref{eqn_betweenness_centrality} is not continuous as defined in Definition \ref{def_continuity}.
\end{proposition}
%%%%%%%%%%%%%%%%%%%%%%%%%%   P   R   O   O   F   %%%%%%%%%%%%%%%%%%%%%%%%%%%%%%%%%% 
\begin{myproofnoname}
The same counter-example used in the proof of Proposition \ref{prop_centrality_unstable} can be used to show failure of continuity. As $\epsilon \rightarrow 0$, we have that $||B||_2 \rightarrow 0$. However, $\left| C_B^G(x) - C_B^H(x) \right| \rightarrow 9$, violating Definition \ref{def_continuity}.
\end{myproofnoname}
%%%%%%%%%%%%%%%%%%%%%%%%%%%%%%%%%%%%%%%%%%%%%%%%%%%%%%%%%%%%%%%%%%%%%

Being not only unstable but discontinuous further hinders practical applicability of $C_B$. Given that $C_{SB}$ captures a similar notion but does so while being stable, thus continuous, makes it an appealing alternative. The numerical experiments in the following section further illustrate how the undesirable structural properties of betweenness centrality translate into lack of robustness when applied to synthetic and real-world data.

%%%%%%%%%%%%%%%%%%%%%%%%%%%%%%%%%%%%%%%%%%%%%%%%%%%%%%%%%%%%%%%%%%%
%%%%%%%%%%%%%%%%   N   U   M   E   R   I   C   A   L        E   X   P   E   R   I   M   E   N   T   S  %%%%%%%%%%%%%%%%%%%
%%%%%%%%%%%%%%%%%%%%%%%%%%%%%%%%%%%%%%%%%%%%%%%%%%%%%%%%%%%%%%%%%%%
\section{Numerical Experiments}
\label{sec_numerical_experiments}

Stability and continuity regulate the behavior of centrality measures in the presence of noise. We empirically validate three facts: the behavior of betweenness centrality in the presence of noise is fundamentally different from the other measures (Section \ref{subsec_robustness_indicators}), continuity and stability encode different robustness properties (Section \ref{subsec_effects_continuity_stability}), and the stable betweenness alternative $C_{SB}$ retains the same centrality notion as the original $C_B$ (Section \ref{sub_sec_ranking_similarity_across_measures}).

For a given node set $V$ of size $n \geq 10$, we define a random network as one where an undirected edge $(x, x')$ belongs to $E$ with probability $q=10/n$. The weight of this edge is randomly picked from a uniform distribution in $[0.5,1.5]$. We consider these weights to be indication of dissimilarities. Notice that the centrality rankings obtained by applying a centrality measure based on dissimilarities -- e.g., closeness -- and one based on similarities -- e.g., degree -- on the same graph are not comparable. Thus, for every random graph we generate a similarity based graph with the same nodes and edges but where the weights are computed as 2 minus the edges in the original dissimilarity graph. In this way, all weights in the similarity graphs are also contained in $[0.5, 1.5]$ and all centrality rankings can be compared. Closeness, betweenness and stable betweenness centralities will be applied to dissimilarity networks while eigenvector and degree centrality will be applied to similarity networks.

As real-world data, we use two networks, one contains information about the air traffic between the most popular airports in Unites States (U.S.) \cite{Colizzaetal07} while the second network records interactions between sectors of the U.S. economy \cite{USinputoutput}. More precisely, in the undirected airport network $G_A = (V_A, E_A, W_A)$, the node set $V_A$ is composed of 25 popular airports in U.S., an edge $(x, x')$ exists between two airports $x, x' \in V_A$ if there is a regularly scheduled flight between them, and the weight of this edge $W_A(x, x')$ is equal to the number of passenger seats -- either occupied or empty -- between both destinations in a given year. The economic network $G_I = (V_I, E_I, W_I)$, contains as nodes the 61 industrial sectors of the economy as defined by the North American Industry Classification System (NAICS). There exists an edge $(x, x') \in E_I$ if part of the output of sector $x$ is used as input to sector $x'$, and the weight $W_I(x, x')$ is given by how much output of $x$ -- in dollars -- is productive input of $x'$. We consider both $W_A(x, x')$ and $W_I(x, x')$ as measures of similarity and use the inverses $1/W_A(x, x')$ and $1/W_I(x, x')$ as weights for the centrality measures that require dissimilarity graphs.

\subsection{Robustness indicators}\label{subsec_robustness_indicators}

%%%%%%%%%%%%%%%   F   I   G   U   R    E   :   R    A    N    D        N   E   T   W   O   R   K   %%%%%%%%%%%%%%%%%%%%%
\begin{figure*}
\centering

\begin{subfigure}{.33\textwidth}
  \centering
  \includegraphics[width=\textwidth]{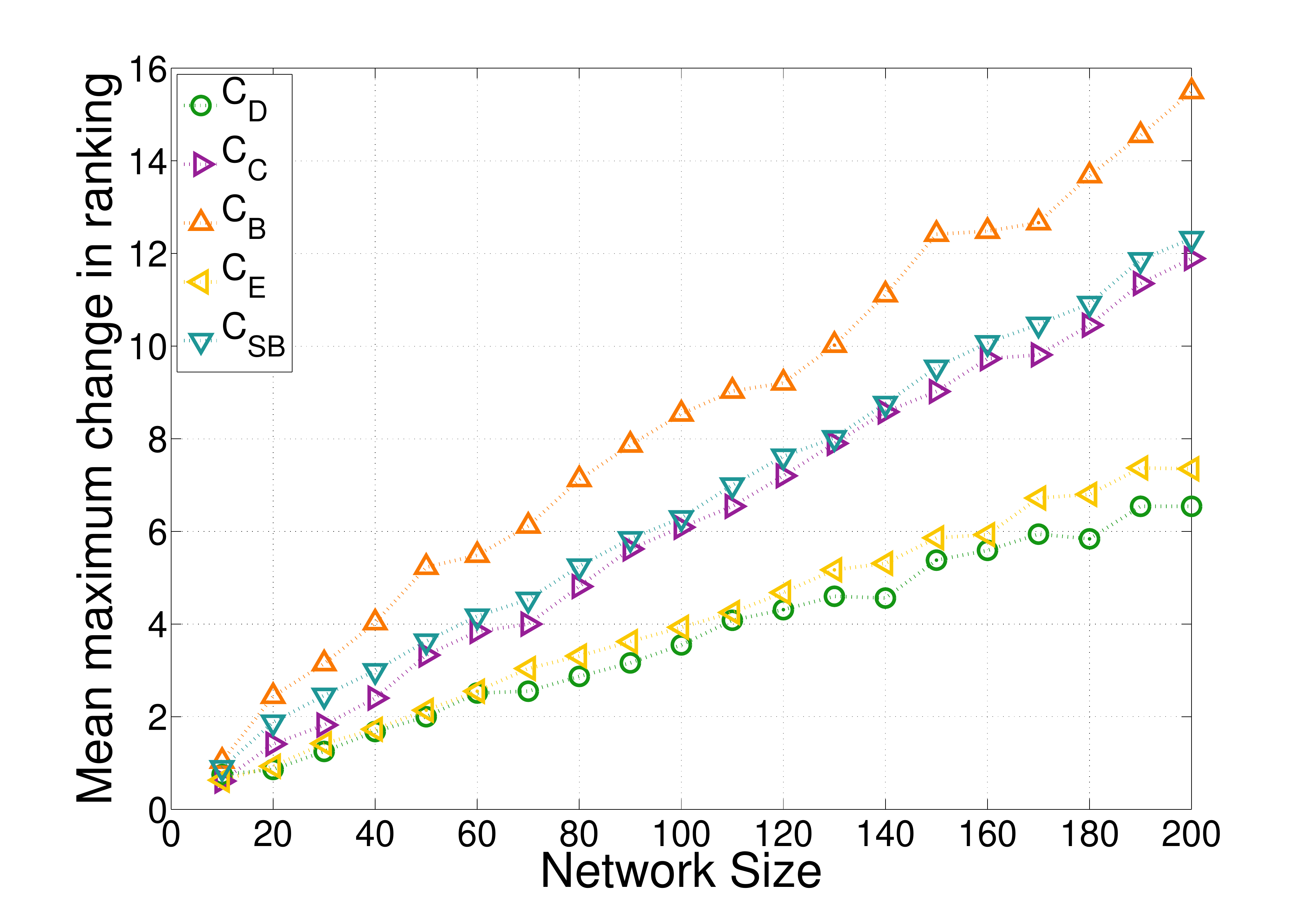}
  \caption{}
  \label{fig:sub7}
\end{subfigure}%
\begin{subfigure}{.33\textwidth}
  \centering
  \includegraphics[width=\textwidth]{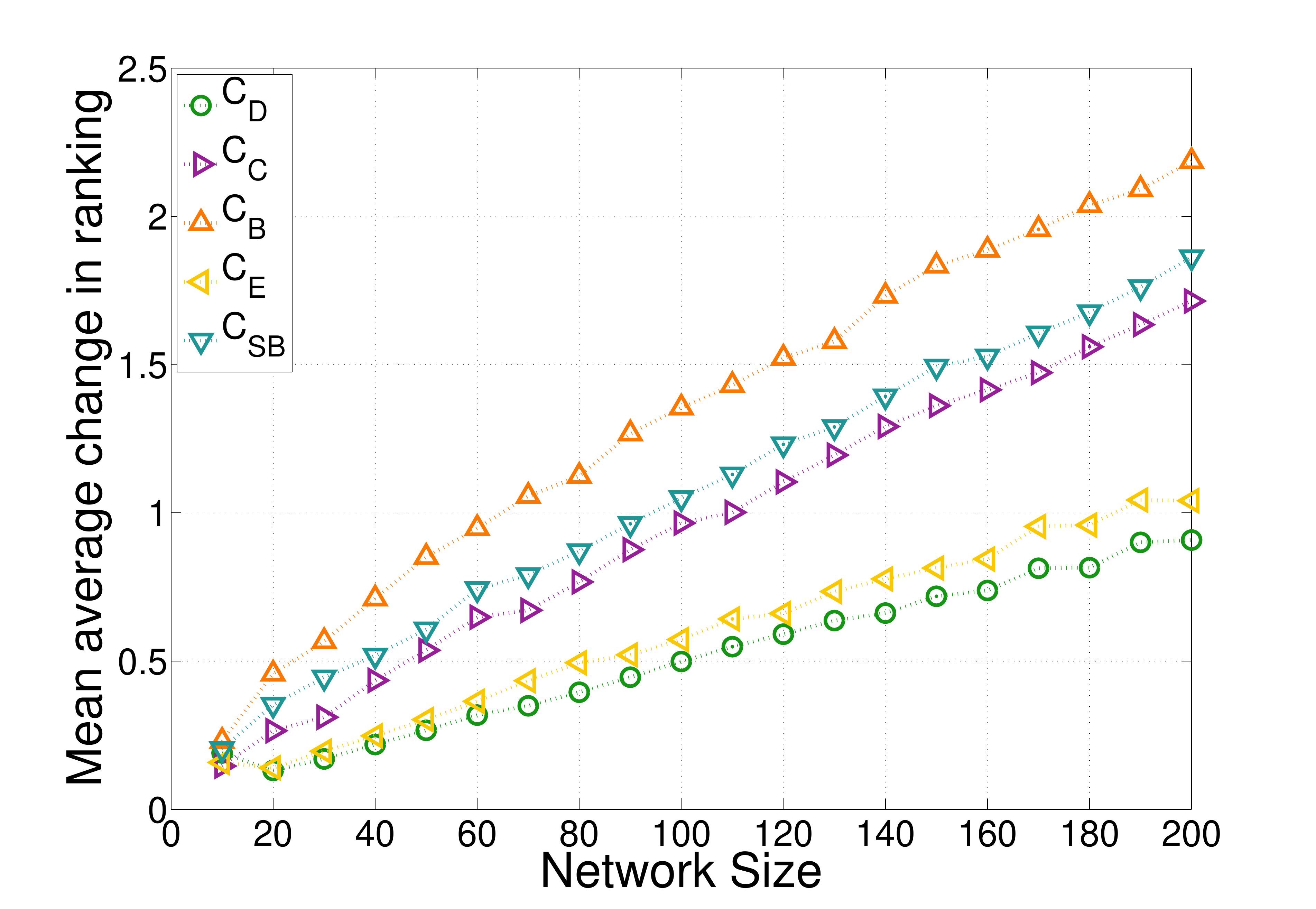}
  \caption{}
  \label{fig:sub8}
\end{subfigure}%
\begin{subfigure}{.33\textwidth}
  \centering
  \includegraphics[width=\textwidth]{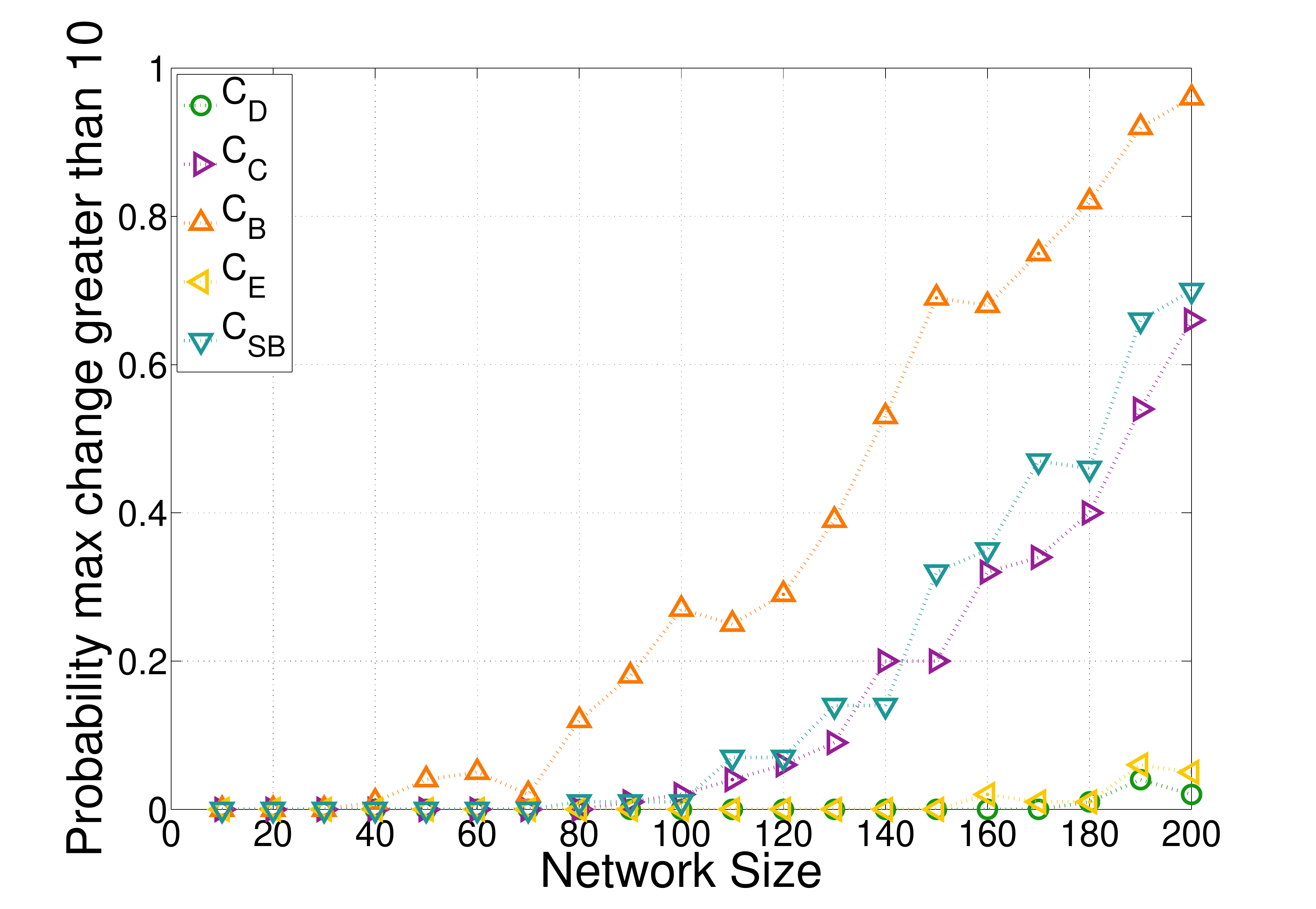}
  \caption{}
  \label{fig:sub9}
\end{subfigure}

\caption{Comparison of stability indicators when type 2 noise ($p_2 = 0.1$, $\delta_2 = 0.1$) is introduced in random networks for all centrality measures: degree (green circle), closeness (purple right triangle), betweenness (orange upwards triangle),  eigenvector (yellow left triangle), and stable betweenness (cyan downwards triangle). (a) Mean of the maximum change recorded when perturbing a random network as a function of network size. (b) Mean of the average node rank change recorded when perturbing a random network as a function of network size. (c) Probability that the maximum change in the ranking exceeds 10 positions as a function of the network size.}
\vspace{-0.1in}
\label{fig_experiments_2}
\end{figure*}
%%%%%%%%%%%%%%%%%%%%%%%%%%%%%%%%%%%%%%%%%%%%%%%%%%%%%%%%%%%%%%%%%%%

We analyze the robustness of the centrality rankings when the random networks are perturbed by random noise. Our specification of random noise has two parameters: the probability of perturbation $p$ and the amplitude of perturbation $\delta$. Given a network, we build a perturbed version of it by modifying every edge weight with probability $p$. The perturbed edge weights are multiplied by a uniform random number in $[1-\delta, 1+\delta]$. In our simulations, we analyze two kinds of noise: type 1 noise has parameters $p_1=1$ and $\delta_1 = 0.01$ while type 2 noise has parameters $p_2=0.1$ and $\delta_2=0.1$. The first noise affects every edge but modifies the weight by a maximum of 1\% whereas the second type of noise affects on average one out of every ten edges but modifies the weight up to 10\%.

For the following experiment, we generate 100 random networks of $n$ nodes, where $n$ varies from 10 to 200 in multiples of 10. We then generate two perturbed versions of each of these networks by applying both types of noises. For every network, we generate a centrality ranking of the nodes, i.e. we sort the nodes in decreasing order of centrality value, and compare it with the centrality ranking of the perturbed versions of that network. We perform this comparison for the rankings output by the four commonly used centrality measures -- degree, closeness, betweenness and eigenvector -- as well as the stable betweenness centrality measure introduced in Section \ref{sec_stable_centrality_measure}.

%%%%%%%%%%%%%%%   F   I   G   U   R    E   :   R    A    N    D        N   E   T   W   O   R   K   %%%%%%%%%%%%%%%%%%%%%
\begin{figure*}
\centering

\begin{subfigure}{.33\textwidth}
  \centering
  \includegraphics[width=\textwidth]{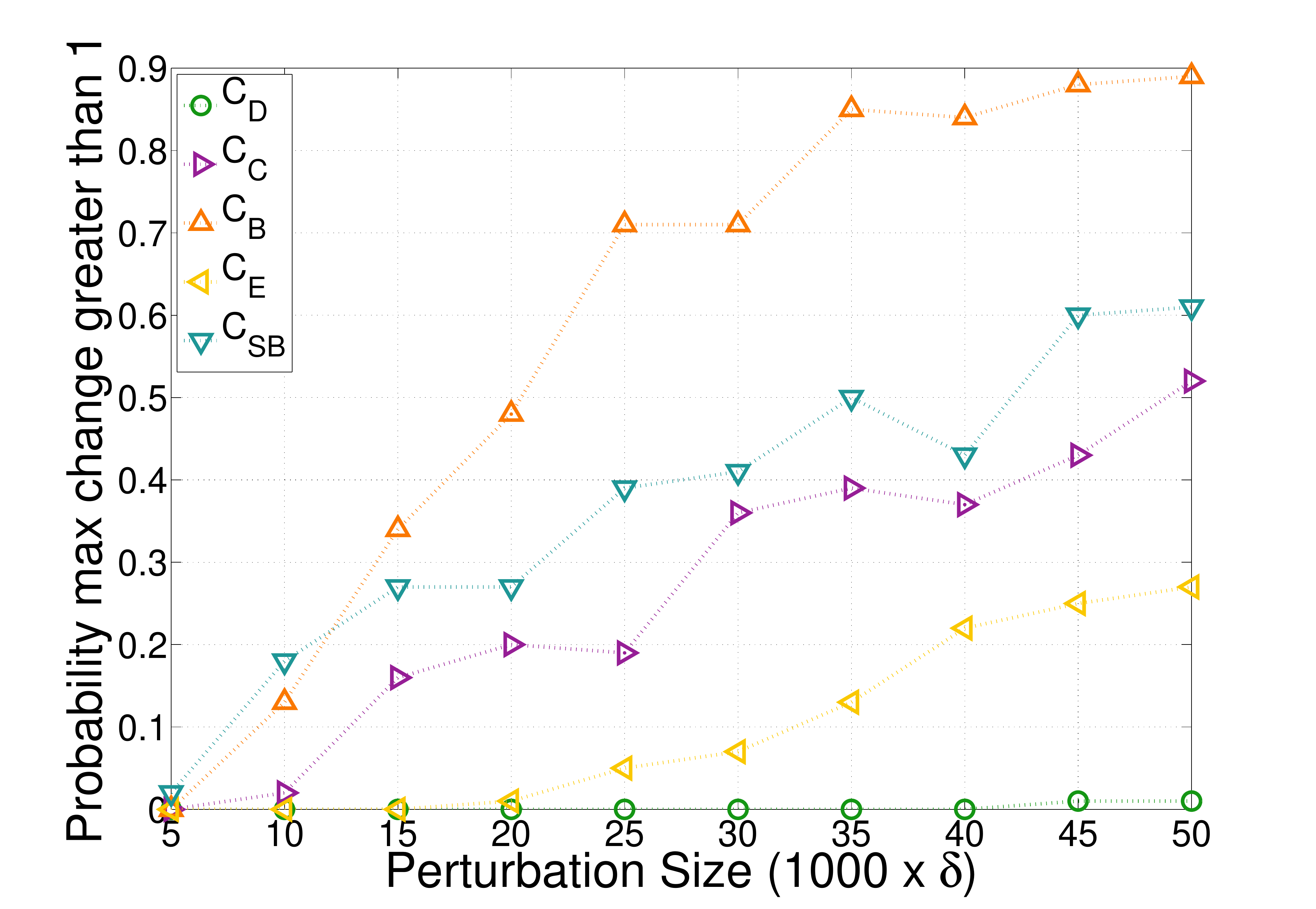}
  \caption{}
  \label{fig:sub1_re}
\end{subfigure}%
\begin{subfigure}{.33\textwidth}
  \centering
  \includegraphics[width=\textwidth]{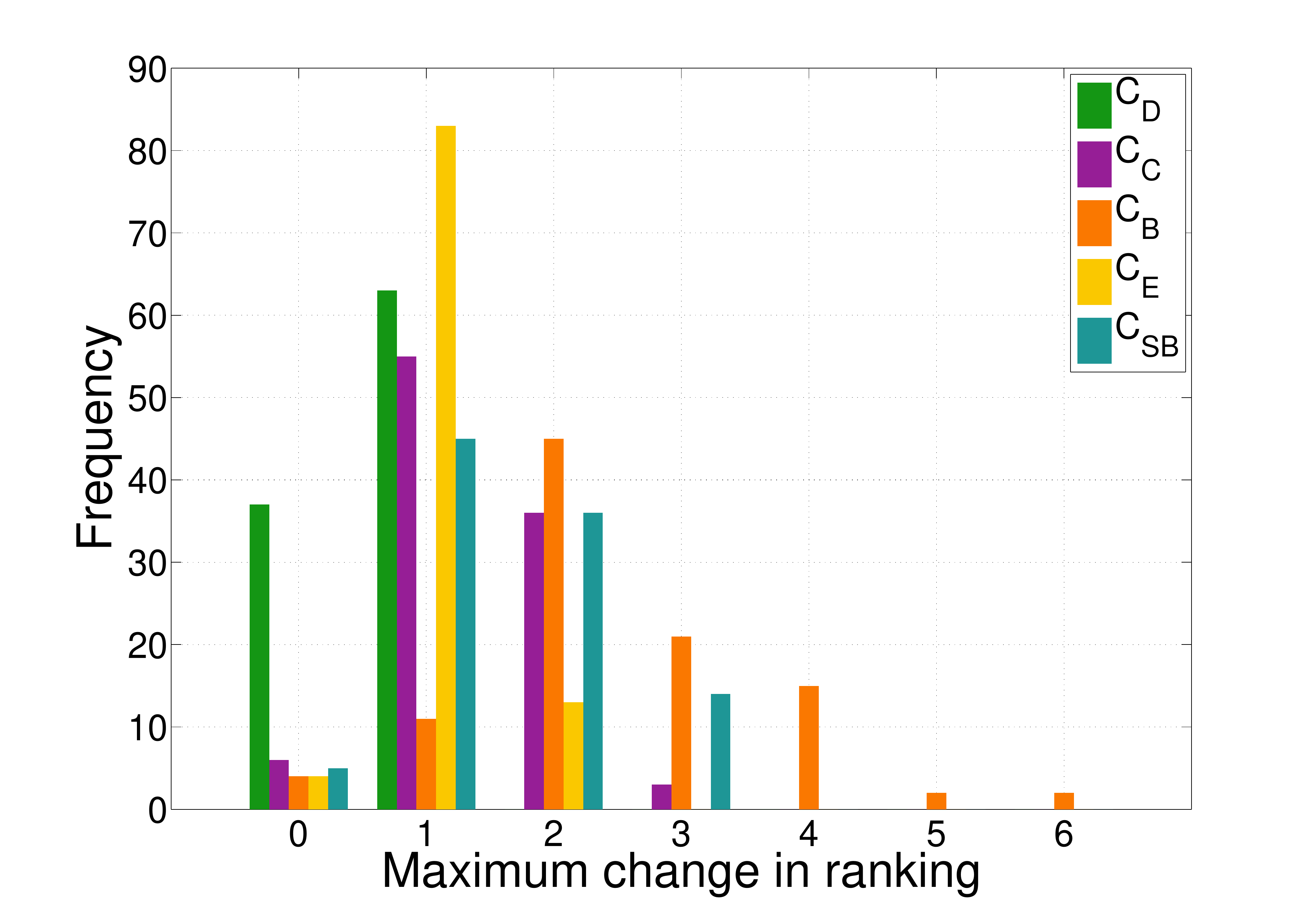}
  \caption{}
  \label{fig:sub2_re}
\end{subfigure}%
\begin{subfigure}{.33\textwidth}
  \centering
  \includegraphics[width=\textwidth]{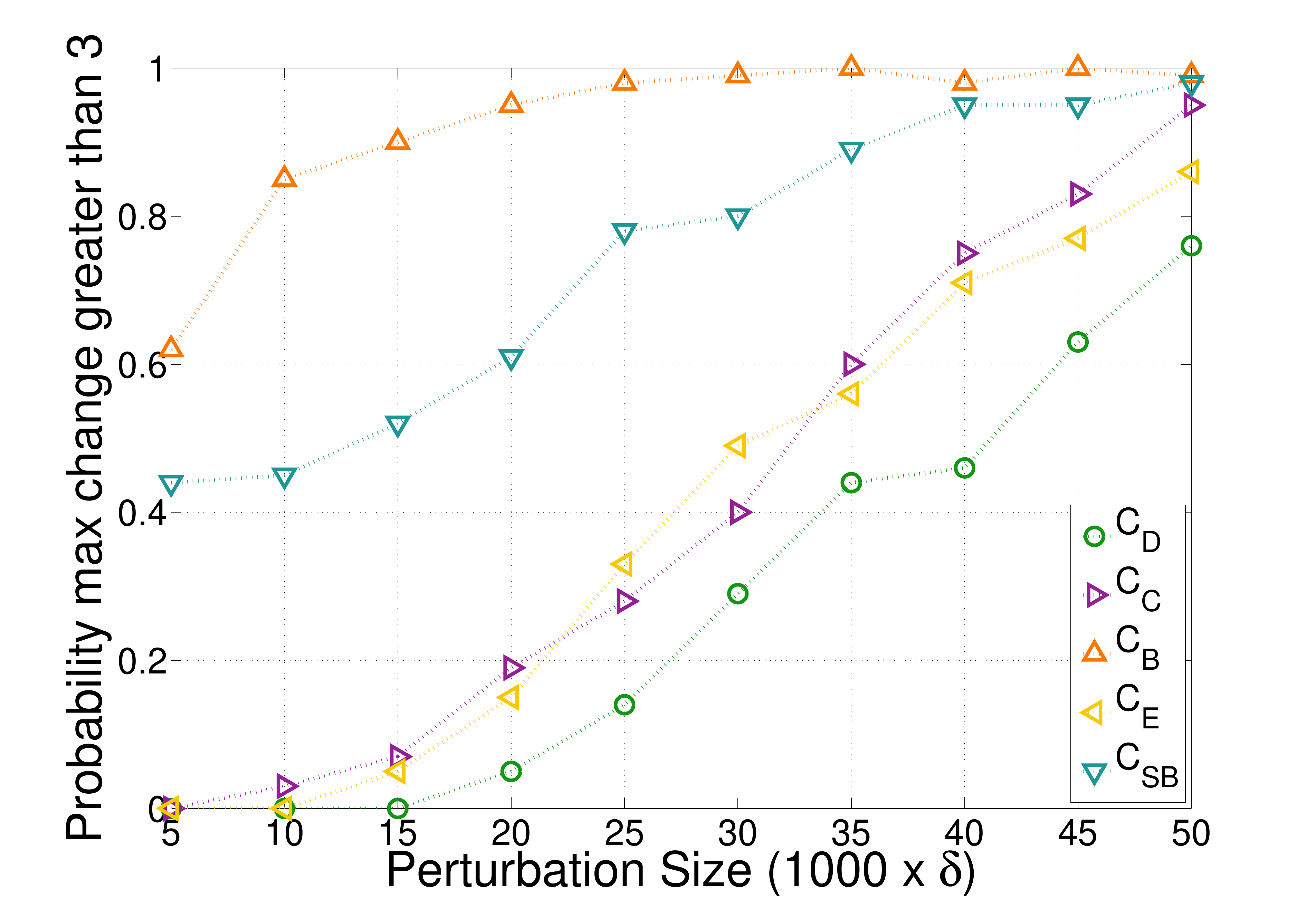}
  \caption{}
  \label{fig:sub3_re}
\end{subfigure}

\caption{Comparison of stability indicators in real-world networks for all centrality measures: degree (green circle), closeness (purple right triangle), betweenness (orange upwards triangle),  eigenvector (yellow left triangle), and stable betweenness (cyan downwards triangle). (a) Probability that the maximum change in the ranking of the airport network $G_A$ exceeds 1 position as a function of the perturbation size. (b) Histogram of the maximum change recorded when perturbing the airport network $G_A$ with $\delta = 0.035$. (c) Probability that the maximum change in the ranking when perturbing the economic network $G'_I$ exceeds 3 positions as a function of the perturbation size.}
\vspace{-0.1in}
\label{fig_experiments_real_world}
\end{figure*}
%%%%%%%%%%%%%%%%%%%%%%%%%%%%%%%%%%%%%%%%%%%%%%%%%%%%%%%%%%%%%%%%%%%

A number of stability indicators are analyzed when perturbing the networks with both types of noise; see Figs. \ref{fig_experiments} and \ref{fig_experiments_2}. For type 1 noise, we begin by analyzing the maximum variation in ranking position experienced by a node when perturbing the network. In Fig. \ref{fig:sub1} we plot the mean of this indicator among the networks analyzed as a function of the network size. For example, for a network with 100 nodes, the type 1 perturbation generates a maximum change of 1.8 positions on average for the $C_{D}$ ranking, 2.6 positions on average for the $C_{C}$ ranking, 5.9 positions on average for the $C_{B}$ ranking, 2.0 positions on average for the $C_{E}$ ranking, and 2.7 positions for the $C_{SB}$ ranking. All measures experience an approximately linear increase of the maximum change with the size of the network, but the rate of increase is fastest for  $C_B$, generating big performance differences between the measures for larger networks. Moreover, the behavior of degree and eigenvector centrality as well as the behavior of closeness and stable betweenness centrality are similar to each other. This is not surprising since they depend upon similar properties of the network. Both closeness and stable betweenness are defined in terms of shortest paths. Also, both degree and eigenvector centrality depend on a notion of neighborhood of each node. For the former, centrality coincides with the graph theoretic notion of neighborhood whereas for the latter, centrality depends on a neighborhood weighted by its influence. In Fig. \ref{fig:sub2} we plot the mean average change when perturbing the network with type 1 noise. I.e., the expected rank variation of any given node in the network. The trend is very similar to the one for maximum variation in ranking. E.g., for a network containing 150 nodes, on average every node experiences a change in 1 position for betweenness centrality while the change is around 0.5 positions for closeness and stable betweenness and 0.3 for degree and eigenvector centralities. Apart from computing the mean rank variations across networks, we are interested in the distributions of these variations for the different centrality measures. Thus, we plot the probability that the maximum change in the ranking generated by a perturbation of type 1 is greater than 3 positions (Fig. \ref{fig:sub3}) and 5 positions (Fig. \ref{fig:sub4}) as a function of the network size. E.g., for networks of 60 nodes, there is a 0.5 probability that the betweenness centrality ranking undergoes a variation greater than 3 positions while this probability is less than 0.1 for all other measures. Moreover, for over 90\% of the networks of 180 nodes, the betweenness centrality ranking undergoes a variation greater than 5 positions when perturbed while this percent is smaller than 10\% for the other measures. To facilitate the understanding of Figs. \ref{fig:sub1}, \ref{fig:sub3}, and \ref{fig:sub4}, in Fig. \ref{fig:sub5} we present the histogram of the maximum change found in the rankings when perturbing a network for the particular case of networks with 150 nodes for all measures. The mean of these histograms correspond to the markers for networks with 150 nodes in Fig. \ref{fig:sub1}. In this way, the mean of the green histogram corresponds to the green circle, the orange histogram to the orange upwards triangle and so on. To relate the histogram with Fig. \ref{fig:sub3}, notice that the green histogram has a frequency of 7 for changes of 4 positions and zero frequency for larger changes. Since we consider 100 sample networks of each size, this translates into a 0.07 probability of observing changes greater than 3 positions for networks of size 150 nodes, which corresponds to the green circle in Fig. \ref{fig:sub3}. The same is true for Fig. \ref{fig:sub4}, but considering changes greater than 5 positions in the histogram. It is immediate that only the orange histogram has a considerable portion of its weight for changes of 6 positions or more, translating into a big difference in probabilities between the orange marker and the rest in Fig. \ref{fig:sub4}. Having a longer tail, the silhouette of the orange $C_B$ histogram is essentially different from the rest. E.g., for one of the studied networks, the $C_B$ ranking presents a change of 19 positions when the perturbation is introduced whereas the largest variation for all other measures combined is of 8 positions. This is an empirical example of instability as shown in Proposition \ref{prop_centrality_unstable}.

Another indicator we analyze is the position where the change in the ranking occurs. A change towards the last positions of the ranking is irrelevant whereas a change varying the positions of the most central nodes carries important implications. In Fig. \ref{fig:sub6}, we plot the probability that the top 5 nodes in the ranking retain their positions after perturbing the network with type 1 noise. Observe that there is no clear trend with the size of the network but probabilities oscillate around different values for different centrality measures. In this way, we can state that for around 75\% to 95\% of the networks there is no change in the top 5 centrality ranking computed with all measures except for betweenness centrality where this percentage falls to around 60\% on average.

The same conclusions can be extracted when perturbing the networks with type 2 noise; see Fig. \ref{fig_experiments_2}. Even though the difference between $C_B$ and the rest of the measures is not as marked as with type 1 noise, it is immediate that betweenness centrality entails the largest maximum change for every network size (Fig. \ref{fig:sub7}) as well as the largest average change (Fig. \ref{fig:sub8}). Also, the probabilities of having changes in the ranking greater than 10 positions is consistently around 0.25 larger in $C_B$ compared to centralities based in shortest paths -- $C_C$ and $C_{SB}$ -- while this probability is negligible for $C_D$ and $C_E$ (Fig. \ref{fig:sub9}).

Similar behaviors in the presence of noise can be observed when analyzing the real-world data; see Fig. \ref{fig_experiments_real_world}. Notice that for graphs $G_A$ and $G_I$, the network size is fixed. Thus, we analyze performance metrics as a function of the magnitude of the perturbation. A perturbation magnitude of $\delta$ implies that every weight in the network is multiplied by a random number in $[1-\delta, 1+\delta]$. For every perturbation level, we generate 100 perturbed networks. In Fig. \ref{fig:sub1_re}, we compute the probability of observing a change in the ranking of more than 1 position as a function of the magnitude of the perturbation. As expected, the probability of observing a change in the network increases with the perturbation magnitude. Moreover, for a fixed magnitude of perturbation, larger probabilities of variations are observed in the rankings generated by $C_B$ compared to those generated by all other measures. E.g., for a perturbation of 0.035, 85\% of the rankings generated by $C_{B}$ presented a change greater than 1 position whereas among the other measures, $C_{SB}$ presented the greater variation with only 50\% of the analyzed networks. To clarify this plot, in Fig. \ref{fig:sub2_re} we present the histogram of maximum changes observed for a perturbation of $\delta=0.035$. For example, all networks analyzed presented either no change or a change of only one position for $C_D$, thus, the corresponding marker for degree centrality in Fig. \ref{fig:sub1_re} is at null probability for $\delta=0.035$. Similarly, only 15 networks out of the 100 analyzed presented either no change or just 1 position change for $C_{SB}$, resulting in the probability of 0.85 for changes greater than one position plotted in Fig. \ref{fig:sub1_re}. As was the case for random networks [cf. Fig. \ref{fig:sub5}], the histogram corresponding to measure $C_B$ presents a longer tail than the rest which is an empirical proof of instability. We applied this same procedure to the second real-world network $G_I$. Notice that $G_I$ is directed, thus, in order to compare all centrality measures including eigenvector centrality, we symmetrized $G_I$ into $G'_I$ by generating undirected edges with weights equal to the mean of the weights in both directions. In Fig. \ref{fig:sub3_re} we plot the probability that $G'_I$ experiences a change of more than 3 positions in the ranking for varying perturbation magnitudes. As expected, this probability is consistently highest for $C_{SB}$, and the difference with other measures is maximized for perturbation of $\delta = 0.02$ and smaller.

\subsection{Effects of continuity and stability}\label{subsec_effects_continuity_stability}

%%%%%%%%%%%%%%%%%%%%%%%%%%%   F   I   G   U   R   E  %%%%%%%%%%%%%%%%%%%%%%%%%%%%%%%%
\begin{figure}
\centering
  \includegraphics[width=0.5 \textwidth, height = 0.35 \textwidth]{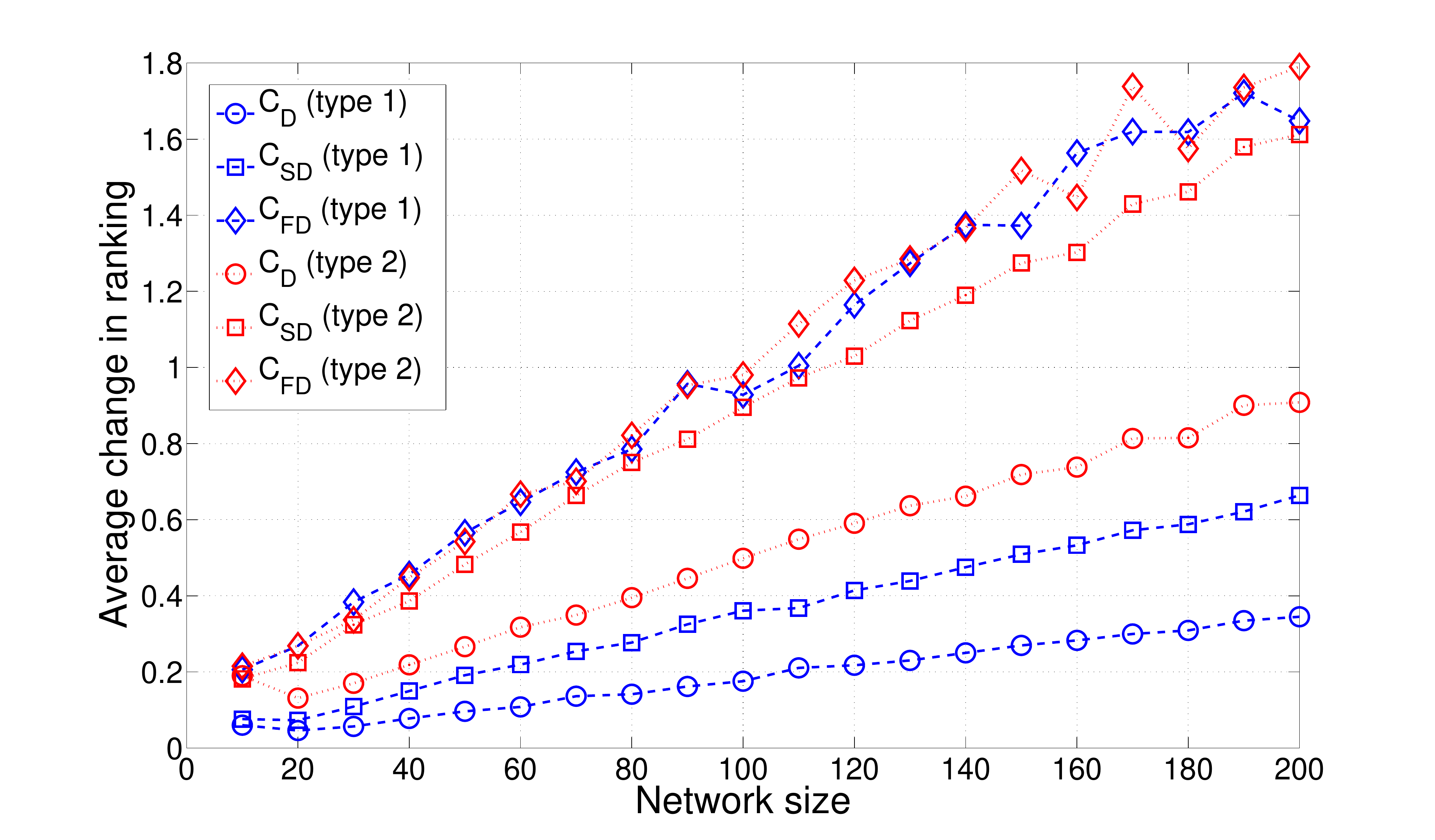}
  \caption{Average change in ranking for degree, degree squared and floor degree centrality measures under two types of noise. For small perturbations (blue), degree squared behaves similar to degree centrality. For larger perturbations (red), degree squared behaves similar to floor degree centrality.}
  \label{fig_degree_centralities_average}
\end{figure}  
%%%%%%%%%%%%%%%%%%%%%%%%%%%%%%%%%%%%%%%%%%%%%%%%%%%%%%%%%%%%%%%%%%%%%

The previous experiment points towards the conclusion that, in practice, stable and continuous centrality measures output centrality rankings with variations which are less meaningful and smaller in magnitude than those obtained with non continuos and non stable measures such as betweenness centrality. Given that betweenness centrality is neither continuous nor stable and the rest of the measures analyzed are both stable and continuous, it is unclear the lack of which property is responsible for the low robustness of betweenness centrality. In order to answer this question, we compare three centrality measures: degree centrality $C_D$ which is both continuous and stable, degree squared centrality $C_{DS}$ as defined in \eqref{eqn_def_sq_degree} which is continuous but not stable, and floor degree centrality $C_{FD}$ which is neither continuous nor stable and we define as follows. For every node $x \in V$ in an arbitrary graph $(V,E,W)$, we have that
\begin{equation}\label{eqn_def_floor_degree_centrality}
C_{FD}(x) := \sum_{x' |(x, x') \in E} \text{floor}\big( W(x, x') \big).
\end{equation}
The fact that $C_{FD}$ is a non-continuous centrality measure is immediate from the discontinuity in the definition of the floor function. In Fig. \ref{fig_degree_centralities_average} we plot the average change in rankings output by the three measures when perturbing networks of different sizes. The results for small noise uniform across edges (type 1) is plotted in blue while the result for larger and sparser noise (type 2) is plotted in red. As expected, degree centrality has the higher robustness followed by degree squared and floor degree being the less robust of the three measures under both types of noise. However, notice that for noise of small magnitude (type 1) the degree squared behaves more similar to degree centrality, showing a robust behavior in the presence of noise. For larger magnitudes of noise (type 2), degree squared centrality has a similar behavior to the unstable floor degree centrality. This points towards the fact that continuity provides robustness under small perturbations while the stronger concept of stability provides robustness for more general perturbations.

\subsection{Ranking similarity across measures}\label{sub_sec_ranking_similarity_across_measures}

%%%%%%%%%%%%%%%%%%%%%%%%%%%   T   A   B   L   E   %%%%%%%%%%%%%%%%%%%%%%%%%%%%%%%%%
\begin{table}
\small
\centering
\caption{Average and maximum variation of centrality ranking across different measures for networks with 100 nodes. The upper triangular part of the table informs the average variation while the lower triangular part informs the maximum variation for the corresponding pair of measures in the rows and columns.}
\begin{tabular}{ c  c  c  c  c  c  c }
  \hline    
  & $C_D$ & $C_C$ & $C_B$ & $C_E$ & $C_{SB}$ & $C_{DS}$ \\  \hline       
   $C_D$ &      0   & 11.3   & 11.6    & 7.3  & 13.1  &  5.3 \\ 
  $C_C$ &  43.8      &   0  & 10.3   & 9.9  & 12.7   & 8.5 \\ 
  $C_B$ &  44.7 &  41.6   &      0   &14.6&    {\bf 4.2}  &  8.3 \\ 
   $C_E$ &  30.0   & 38.9  & 55.5       &  0   &16.6  &  8.5 \\ 
  $C_{SB}$ &  51.1 &  51.3 &   {\bf 18.9} &  61.5 &        0  & 10.0 \\ 
   $C_{DS}$ &  22.3 &  34.4  & 34.3 &  33.7 & 42.4    &     0 \\ \hline
\end{tabular}
\label{table_avg_max_variation_centrality_measures}
\end{table}
%%%%%%%%%%%%%%%%%%%%%%%%%%%%%%%%%%%%%%%%%%%%%%%%%%%%%%%%%%%%%%%%%%%%

In order to compare the centrality rankings across different measures, we pick 100 random networks of size 100 nodes and compute the average and maximum change for a pair of rankings output by different measures; see Table \ref{table_avg_max_variation_centrality_measures}. E.g., in this 100 samples the mean average ranking variation of nodes ranked by the degree $C_D$ and the eigenvector $C_E$ centralities is 7.3 positions. Moreover, the mean maximum variation between two given rankings output by the betweenness $C_B$ and the closeness $C_C$ centrality is 41.6 positions. Notice that the smallest variations -- both in average and maximum -- are achieved when comparing the rankings of the betweenness $C_B$ and the stable betweenness $C_{SB}$ centrality measures. This is empirical proof that both measures encode a similar centrality concept, as was our objective when defining stable betweenness centrality in Section \ref{sec_stable_centrality_measure}. Further observe that the variations between these two rankings are even smaller than the ones between degree $C_D$ and squared degree $C_{DS}$ centrality, two measures with closely related definitions [cf. \eqref{def_degree_centrality} and \eqref{eqn_def_sq_degree}].

To complete the analysis, we use the economic network $G_I$ to illustrate the fact that the centrality concept in the proposed measure $C_{SB}$ closely resembles the one in the traditional betweenness centrality $C_B$; see Table \ref{table_io_betweenness_stable_betweenness}. To avoid introducing artifacts through symmetrization, we consider the original network $G_I$ instead of the symmetrized version $G'_I$, hence, the eigenvector centrality is not informed. Stable betweenness centrality $C_{SB}$ provides the ranking closer to the one output by $C_B$. Both measures share the top 3 economic sectors and 4 out of 5 sectors in the top 5. In contrast, none of the other measures -- closeness, out-degree, and in-degree -- contain the three sectors preferred by $C_B$ in their top 5 ranking.

%%%%%%%%%%%%%%%%%%%%%%%%%%%   T   A   B   L   E   %%%%%%%%%%%%%%%%%%%%%%%%%%%%%%%%%
\begin{table*}
\small
\centering
\caption{Comparison of the centrality rankings for the economic network $G_I$. The ranking output by $C_{SB}$ is the closest one to the $C_B$ ranking. $C_{B}$ and $C_{SB}$ share the top 3 ranking while these three economic sectors are not contained in the top 5 ranking of any other measure. }
\begin{tabular}{ l  l  l l l l}
  \hline    
  Rank & $C_B$ & $C_{SB}$ & $C_C$ & $C_{OD}$ & $C_{ID}$ \\  \hline       
   {\bf 1} & {\bf Real estate} &   {\bf Professional serv.}  &  {\bf Professional serv.}    & {\bf Professional serv.} &  Food \& Beverage \\ 
   {\bf 2} & {\bf Construction}   & {\bf Real Estate} &    Oil and gas extraction &   {\bf Real Estate} &   {\bf Real Estate}  \\ 
   {\bf 3} & {\bf Professional serv.}   &  {\bf Construction}  &  Petroleum products  &   Oil and gas extraction &   Petroleum products \\ 
   4 & Wholesale trade     & Petroleum products  &  Administrative serv. &   FR banks, credits &   Chemical products \\  
   5 & FR banks, credits    & FR banks, credits  &  {\bf Real estate}  &  Administrative services  &   {\bf Construction} \\  \hline
\end{tabular}
\label{table_io_betweenness_stable_betweenness}
\end{table*}
%%%%%%%%%%%%%%%%%%%%%%%%%%%%%%%%%%%%%%%%%%%%%%%%%%%%%%%%%%%%%%%%%%%%

%%%%%%%%%%%%%%%%%%%%%%%%%%%%%%%%%%%%%%%%%%%%%%%%%%%%%%%%%%%%%%%%%%%%%

%%%%%%%%%%%%%%%%%%%%%%%%%%%%%%%%%%%%%%%%%%%%%%%%%%%%%%%%%%%%%%%%%%%%%
%%%%%%%%%%%%%%%%%%%%%%%      C O N C L U S I O N        %%%%%%%%%%%%%%%%%%%%%%%%%%%%%%%%
%%%%%%%%%%%%%%%%%%%%%%%%%%%%%%%%%%%%%%%%%%%%%%%%%%%%%%%%%%%%%%%%%%%%%
\section{Conclusion}\label{sec:conclusion}

Stability, as a formal characterization of the robustness of node centrality measures, was introduced. The most commonly used centrality measures were shown to be stable with the exception of betweenness centrality, thus, a stable alternative definition was proposed. A milder continuity property was introduced and betweenness centrality was shown not to be continuous. We illustrated the stability difference between betweenness centrality and the rest of the measures by studying indicators in both random and real-world networks. Moreover, by proposing alternative definitions of degree centrality, the practical differences between continuity and stability were exemplified. Finally, by comparing the centrality rankings output by different measures, it was shown that stable betweenness preserves the centrality notion encoded in traditional betweenness but has the additional practical advantage of stability. 

%%%%%%%%%%%%%%%%%%%%%%%%%%%%%%%%%%%%%%%%%%%%%%%%%%%%%%%%%%%%%%%%%%%%%

%\vfill\pagebreak

%****************************************************************************************************************************************************%
%-----------------------------------------------------------------------------------------------------------------------------------------------------------------------------------------%
%%%%%%%%%%%%%%%%%%%%%                    S E C T I O N                      %%%%%%%%%%%%%%%%%%%%%%%%%%%%%%
%%%%%%%%%%%%%%%%%%%%%%%      R E F E R E N C E S         %%%%%%%%%%%%%%%%%%%%%%%%%%%%%%%%
%-----------------------------------------------------------------------------------------------------------------------------------------------------------------------------------------%
%****************************************************************************************************************************************************%
\urlstyle{same}
\bibliographystyle{IEEEtran}
\bibliography{centrality_biblio.bib}
%****************************************************************************************************************************************************%

\end{document}